\documentclass[prb,aps,reprint,showpacs,groupedaddress]{revtex4-1}
\usepackage{graphicx}
\usepackage{amsmath}
\usepackage{titlesec} 
\usepackage{bbm}
\usepackage{romannum}
\titlespacing{\section}{10pt}{10pt}{10pt}
\usepackage{subfigure}

\begin{document}

\abovedisplayskip=2mm plus 3pt minus 9pt
\abovedisplayshortskip=0pt plus 3pt
\belowdisplayskip=2mm plus 3pt minus 9pt
\belowdisplayshortskip=7pt plus 3pt minus 4pt

\abovedisplayskip=2mm
\belowdisplayskip=2mm
\belowdisplayshortskip=10mm
\abovedisplayshortskip=10mm
\pagenumbering{arabic}

\title{Josephson current through a quantum dot coupled to a molecular magnet}

\author{P. Stadler}
\affiliation{Fachbereich Physik, Universit\"at Konstanz, D-78457 Konstanz, Germany}
\author{C. Holmqvist}
\affiliation{Fachbereich Physik, Universit\"at Konstanz, D-78457 Konstanz, Germany}
\author{W. Belzig}
\affiliation{Fachbereich Physik, Universit\"at Konstanz, D-78457 Konstanz, Germany}

\date{\today}
\begin{abstract}
Josephson currents are carried by sharp Andreev states 
within the superconducting energy gap. We theoretically study the electronic
transport of a magnetically tunable nanoscale junction consisting of a quantum dot connected to two superconducting leads and coupled to the spin of a molecular magnet.  
The exchange interaction between the molecular magnet and the quantum dot modifies the
Andreev states due to a spin-dependent renormalization of the quantum dot's energy level and the induction of spin-flips.  A magnetic field applied to the central region of the quantum dot and the molecular magnet further tunes the Josephson current and starts a precession of the molecular magnet's spin. We use a 
non-equilibrium Green's function approach to evaluate the transport properties of
the junction. Our calculations reveal that the energy level of the dot, the magnetic field and the exchange interaction between the molecular magnet and the electrons occupying the energy level of the quantum dot can trigger transitions from a  
$ 0 $ to a $ \pi $ state of the Josephson junction. The redistribution of the occupied states induced by the magnetic field strongly modifies the current-phase relation. The critical current exhibits a sharp increase as a function of either the energy level of the dot, the magnetic field or the exchange interaction.
\end{abstract}

\maketitle

\section{Introduction}
Molecular spintronics combines the two fields molecular electronics and spintronics. \cite{Rocha:2005ep} In spintronics, the electron spin is used as the degree of freedom in which information is encoded. \cite{Wolf:2001fu} Molecular electronics investigates the electrical and thermal properties of molecules and aims to build devices composed of single molecules or ensembles of molecules. In molecular spintronics, the spin of molecules is used to manipulate the spin and charge transport. In particular, molecular magnets are interesting as basic building blocks for electronic devices \cite{Bogani:2008tc,MolecularSpintronic:2006de,Rocha:2005ep} and for quantum computing. \cite{Leuenberger:478346,Timm:2012cf} These molecules have a permanent magnetisation due the their anisotropy barrier as well as long coherence times \cite{Ardavan:2007ci} that facilitate further quantum-mechanical phenomena such as interference \cite{Wernsdorfer:1999tx,Wernsdorfer:2002jx} and quantum tunnelling of the magnetisation. \cite{Bogani:2008tc,Christou:2000vo,Gatteschi:2003uu}

Experimentally, the transport properties of different kinds of junctions containing magnetic molecules have been extensively studied in three-terminal devices. \cite{Grose:2008ev,Jo:2006ck,Heersche:2006cs,Henderson:2007vi,Haque:2011we,Zyazin:2012gc,Burzuri:2012gb} Current measurements through molecular magnets allows to identify the magnetic states and directly observe the magnetic anisotropy and the orientation of the easy axis. \cite{Burzuri:2012gb} These magnetic states of the molecule have been proposed to enable quantum computing. \cite{Leuenberger:478346} An alternative way to probe the properties of molecular magnets is to deposit the molecules on carbon nanotubes\cite{Urdampilleta:2011ii} or graphene layers. \cite{Candini:2011ul} The presence of the magnetic molecule modifies the transport through the junction and oppositely, the tunnelling electrons modify the magnetisation of the molecule and can reverse the magnetisation. \cite{Candini:2011ul} Other experiments used superconducting electrodes with the advantage that the heat losses in these devices disappear.  The proximity-induced superconductivity and the accompanied Andreev reflections modify the transport properties through the molecule. 
\cite{Kasumov:2005dn,Winkelmann:2009vc} In an Andreev reflection process, incoming electron-(hole-)like quasiparticles with energies lying within the superconducting gap are retroreflected as hole-(electron-)like quasiparticles at a normal-superconductor interface. In a junction consisting of two superconducting leads coupled over a non-superconducting region, the Andreev reflected electron- and hole-like quasiparticles at the left and right interfaces form Andreev levels which carry the Josephson current. Josephson junctions offer the possibility for applications in superconducting electronics, as well as quantum information and computing. \cite{Clarke:2008gi,Wendin:2007da} The Andreev levels of a Josephson junction can in principle be used as a two level qubit. \cite{Zazunov:2003jm,Zazunov:2005ec} In Ref. [\onlinecite{Chtchelkatchev:2003ji}], an Andreev level qubit with spin orbit coupling was discussed as building block to perform quantum computations using the spin degree of freedom in order to manipulate the Andreev states. The manipulation of the Josephson current by adding quasiparticles to the Andreev states and the resulting suppression of the Josephson current were measured in Ref. [\onlinecite{Zgirski:2011dx}]. An alternative way to manipulate the Andreev states of a quantum point contact in the presence of a magnetic scatterer was studied in Ref. [\onlinecite{Michelsen:2008wv}]. The presence of a molecular magnet in a Josephson junction offers a further possibility to tune the current and manipulate the Andreev states. These states are experimentally accessible and have been observed in Ref. [\onlinecite{Pillet:2010ds}]. By comparison of the energy of the Andreev states with the experimental data it is also possible to extract information about the parameters affecting the current in the constriction. 

Besides the spectroscopy of the Andreev states, the measurement of the Josephson current reveals detailed information about the internal structure of the junction. Direct measurement of the current-phase relation in superconducting atomic contacts have been performed in Ref. [\onlinecite{DellaRocca:2007ua}]. The shape of the current-phase relation strongly depends on the details of the contact between the electrodes and is important for  applications in superconducting devices. The state of the system can change from a $ 0 $ to a $ \pi $ state in which the current changes sign. This transition to the $ \pi $ state was proposed in Ref. [\onlinecite{Bulaevskii:1977uj}] as a result of tunnelling through magnetic impurities. Experimentally, this transition was measured in a junction consisting of two superconducting leads coupled over a quantum dot in the Couloumb blockade regime which was occupied with a odd number of electrons. \cite{vanDam:2006fj} The singlet wave function of the Cooper pairs experiences a phase shift of $ \pi $ due to coherent cotunnelling processes, leading to the reversal of the supercurrent. The $ 0 $ to $ \pi $ transition has been studied in many other kinds of Josephson junctions, such as ferromagnetic heterostructures,\cite{Ryazanov:2001jp,Kontos:2002hj} and theoretically analysed in magnetic junctions associated with a molecular magnet in Refs. [\onlinecite{Benjamin:2007fz,Lee:2008ka,Sadovskyy:2011ge,Teber:2010bo,Holmqvist:2011bv,Holmqvist:2012jz,Holmqvist:2013uj,Nussinov:2005id,Zhu:2004dn}].

In this paper, we study the Josephson current through a quantum dot which is coupled to a molecular magnet. We focus on the limit of low temperature ($ T\rightarrow 0$) and on the regime of negligible Coulomb interaction. The magnetic moment of the molecular magnet is assumed to be large enough to allow for a classical treatment of the molecular magnet's magnetisation. Furthermore, we assume that the molecule has an isotropic magnetisation and may, for instance, be a fullerene molecule doped with a magnetic impurity. \cite{Kasumov:2005dn,Roch:2011ht} The spin of the magnetic molecule then interacts with the electrons occupying the quantum dot via the exchange interaction.
In Refs. [\onlinecite{Benjamin:2007fz}] and [\onlinecite{Lee:2008ka}], the Josephson current through an isotropic magnetic molecule was studied in the Kondo regime and in the regime of negligible Coulomb interaction, respectively. In comparison to the work in Ref. [\onlinecite{Benjamin:2007fz}], in this paper a magnetic field is applied to the central region consisting of the quantum dot and the molecular magnet, whose magnetisation then precesses with the Larmor frequency, $ \omega_L $. The magnetic field and a gate voltage applied to the quantum dot introduce additional parameters for manipulating the Andreev states and tuning the current through the junction. We mainly focus on the dependence of the Josephson current on the energy level of the dot and the magnetic field. The manipulation of the Andreev states by the molecular magnet is studied and the parameter range enabling the junction to be in a $ 0 $ or $ \pi $ state can be determined by the critical current. 

The outline of the paper is as follows. In Sec. \ref{sec:model}, we introduce the model Hamiltonian of the junction and describe the terms including the effect of the molecular magnet. The following Sec. \ref{sec:approach} is concerned with the approach used to determine the transport properties of the junction. The results, starting with the density of states of the quantum dot and the Andreev states, are presented in Sec. \ref{sec:results}. This section also includes a discussion about the current-phase relation and the critical current.  We conclude with a summary of the results in Sec. \ref{sec: conclusions}. 

\section{Model}\label{sec:model}
\begin{figure}[!b]
\begin{centering}
\subfigure{
\includegraphics[width=1\columnwidth]{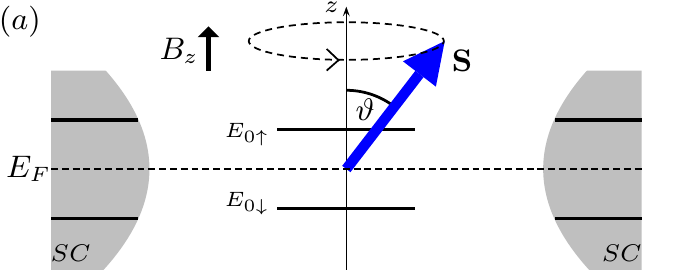}}
\subfigure{
\includegraphics[width=1\columnwidth]{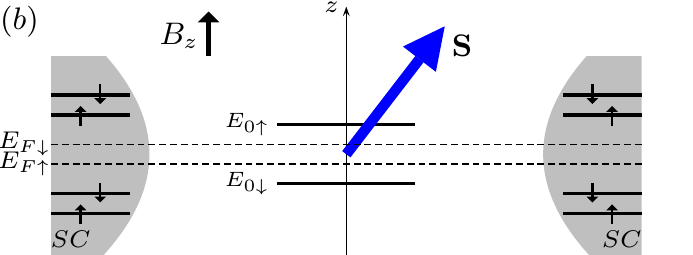}}
\caption{Junction with two superconducting leads and a quantum dot coupled to the spin $\textbf{S}$ of a molecular magnet. In the laboratory frame (panel (a)), the magnetic field forming an angle of $\vartheta$ with the spin, starts a precession of the spin and splits the energy level of the quantum dot into $E_{0\uparrow}$ and $E_{0\downarrow}$. In the frame of the rotating spin (panel (b)), the Fermi energies of the spin-up (spin-down) electrons and the quasiparticle states in the leads are shifted by $-(+)\omega_L/2$. The transformation cancels the Zeeman splitting of the quantum-dot energy level induced by the magnetic field.} 
\label{LFRF}
\end{centering}
\end{figure}

The junction, which is depicted in Fig. \ref{LFRF} (a), consists of two superconducting leads coupled to a quantum dot in the presence of a molecular magnet. A magnetic field applied along the $ z $ axis induces a Zeeman splitting of the quantum dot's energy level as well as a precession of the spin of the molecular magnet. The Hamiltonian of the junction is written as
\begin{equation}
H(t) = \sum_{\alpha=L,R}\left( H_{\alpha} + H_{T\alpha} \right) + H_D + H_S(t) + H_{SD}(t),
\label{SQDSNMFH} 
\end{equation} 
\noindent where the index $ \alpha $ corresponds to the left ($L$) or right ($R$) side of the junction. 
The left and right superconducting leads are described by the BCS Hamiltonian $ H_{\alpha}= \sum_{k_{\alpha}\sigma} \xi_{k_{\alpha}} c^{\dagger}_{k_{\alpha}\sigma}c_{k_{\alpha}\sigma} + \Delta_{\alpha} c^{\dagger}_{k_{\alpha}\uparrow}c^{\dagger}_{-k_{\alpha}\downarrow}+\Delta_{\alpha}^{\dagger} c_{-k_{\alpha}\downarrow} c_{k_{\alpha}\uparrow} $ with $ \sigma = (\uparrow,\downarrow) = \pm 1 $ and the dispersion $ \xi_{k} = \hbar^2k^2/(2m) - \mu $. The order parameter is given by $ \Delta_{\alpha} = \vert \Delta_{\alpha} \vert e^{i\varphi_{\alpha}} $ and for symmetry reasons we assume that $ \varphi_{R,L} = \pm \varphi/2$. In general, the temperature and magnetic field dependence of the order parameter must be taken into account in a self-consistent way. Since we restrict our discussion to the limit $ T \rightarrow 0 $, we neglect the temperature dependence and for simplicity we assume that the applied magnetic field does not fundamentally affect the superconducting leads. Quasiparticles in the leads with momentum $ k $ and spin $ \sigma $ are created and annihilated by the operators $ c^{\dagger}_{k\sigma} $ and $ c_{k\sigma} $. The Hamiltonian describing the tunnelling between the dot and the superconducting leads is written as $H_{T\alpha}  = \sum_{k_{\alpha}\sigma} d^{\dagger}_{\sigma} V_{d_{\sigma},k_{\alpha}} c_{k_{\alpha}\sigma} + c^{\dagger}_{k_{\alpha}\sigma} V^{\dagger}_{d_{\sigma},k_{\alpha}} d_{\sigma} $. The hopping $ V_{d_{\sigma},k_{\alpha}} $ describes the coupling between the dot and the leads and is assumed to be independent of energy. The operators acting on the dot correspond to $ d^{\dagger}_{\sigma} $ and $ d_{\sigma} $. The Hamiltonian of the dot $ H_D $ and the Zeeman energy are written as $ H_D = \sum_{\sigma} (E_0+\sigma \mu_B B_z )d^{\dagger}_{\sigma} d_{\sigma} $ and $  H_{S}(t) = -\boldsymbol{\mu}_{M}(t) \textbf{B} $, with the magnetic field $ \boldsymbol{B} = (0,0,B_z) $ and the magnetic moment of the molecular magnet $  \boldsymbol{\mu}_{M}(t) $. The magnetic moment is related to the spin of the molecular magnet via  $ \boldsymbol{\mu}_{M}(t) =-\gamma \textbf{S}(t) $ with the gyromagnetic ratio  $ \gamma =  g_Me/(2m) $. The Land\'{e} factor of the molecular magnet is denoted by  $ g_M $ and has in principle to be determined by comparison with experiments. In the following, we assume that the Land\'{e} factor of the molecular magnet equals to that of free electrons and $g_M = 2$.\cite{Tserkovnyak:2005zz} Due to the magnetic field, the magnetisation of the molecular magnet precesses with the Larmor frequency. We assume that the motion of the spin is undamped which can be achieved by dc and rf fields. \cite{Kittel:1948ur,Bell:2008gh}  The equation of motion is then given by $ \partial{\textbf{S}}/\partial t = -\gamma \textbf{S} \times \textbf{B} $. The solution of this equation is $
\textbf{S}(t) = S \left(\mathrm{cos}(\omega_L t) \mathrm{sin}(\vartheta)\textbf{e}_x + \mathrm{sin}(\omega_L t) \mathrm{sin}(\vartheta) \textbf{e}_y + \mathrm{cos}(\vartheta) \textbf{e}_z \right) $, with the magnitude of the spin $ \vert \textbf{S} \vert = S $ and the Larmor frequency $ \omega_L = \gamma B_z  $. The exchange interaction between the molecular magnet and the quantum dot is described by $ H_{SD}(t) = \frac{1}{2}\sum_{\sigma\sigma^{\prime}} V_s d_{\sigma}^{\dagger} (\textbf{S}(t) \boldsymbol{\sigma})_{\sigma\sigma^{\prime}}d_{\sigma^{\prime}}$  with the coupling $ V_s $ between the spin and the quantum dot and the Pauli matrices $ \boldsymbol{\sigma} = (\sigma_x,\sigma_y,\sigma_z) $. The term can be transformed into
$
H_{SD}(t)= \sum_{\sigma} \sigma v_s\mathrm{cos}(\vartheta)d^{\dagger}_{\sigma} d_{\sigma} +v_s \mathrm{sin}(\vartheta) e^{-i\omega_Lt} d^{\dagger}_{\uparrow}d_{\downarrow}+v_s \mathrm{sin}(\vartheta)e^{i\omega_Lt}d^{\dagger}_{\downarrow}d_{\uparrow}
$ with $ v_s = SV_s/2 $. The first term induces a spin-dependent shift of the energy levels of the dot while the second and third terms account for the spin flip of the electrons occupying the dot. Since the magnetic field enters in the Hamiltonian of the dot and the exchange interaction, we can rewrite the Hamiltonian of the dot in terms of the Larmor frequency of the molecular magnet as $ H_D = \sum_{\sigma}\left(E_0+\sigma \frac{\omega_L}{2}\right)d^{\dagger}_{\sigma}d_{\sigma} $.

\section{Approach}\label{sec:approach}
The transport properties of the system are described by a nonequilibrium Green's function approach.\cite{Eilenberger:1968bb,Larkin:1969wa,Serene:1983vc} In order to simplify the evaluation of the Green's functions, we perform a unitary transformation to the rotating frame of the molecular magnet's spin, since in this frame the Hamiltonian is time independent.
The state vector transforms according to $ \vert \tilde{\Psi} \rangle = U^{\dagger} \vert \tilde{\Psi} \rangle $ and the Hamiltonian in the rotating frame can be written as $ \bar{H} = U^{\dagger} H U + i (\partial_t U^{\dagger}) U $ with the unitary transformation operator
\begin{equation}
U(t)\mathord=\mathrm{exp}\hspace{-0.1cm}\left[{\mathord-i\frac{\omega_L}{2}t\hspace{-0.1cm}\sum_{\alpha k_{\alpha}}\left(c^{\dagger}_{k_{\alpha}\uparrow} c_{k_{\alpha}\uparrow}\mathord- c^{\dagger}_{k_{\alpha}\downarrow} c_{k_{\alpha}\downarrow}\mathord+d^{\dagger}_{\uparrow}d_{\uparrow}\mathord-d^{\dagger}_{\downarrow}d_{\downarrow}\right)}\hspace{-0.1cm}\right]\hspace{-0.1cm}.
\label{UT}
\end{equation}
In the rotating frame, the Hamilton operator \eqref{SQDSNMFH} is given by 
$ \bar{H} \mathord= \sum_{\alpha}( \bar{H}_{\alpha}\mathord+ \bar{H}_{T\alpha} ) \mathord+ \bar{H}_D \mathord+ \bar{H}_S \mathord+ \bar{H}_{SD} 
$. 
The transformation results in a spin-dependent shift of the quasiparticles' energies in the leads and the quantum dot. The Hamiltonian of the leads reduces to
$
\bar{H}_{\alpha} \mathord= \sum_{k_{\alpha}\sigma} (\xi_{k_{\alpha}\sigma}\mathord-\sigma\frac{\omega_L}{2})  c^{\dagger}_{k_{\alpha}\sigma}c_{k_{\alpha}\sigma} \mathord+ \Delta c^{\dagger}_{k_{\alpha}\uparrow}c^{\dagger}_{-k_{\alpha}\downarrow}+\Delta^{\dagger} c_{-k_{\alpha}\downarrow} c_{k_{\alpha}\uparrow}.
$
Due to the spin-dependent energy shift, the Zeeman energy in the Hamiltonian of the dot is cancelled by the transformation such that the Hamiltonian of the dot is given by $ \bar{H}_D = \sum_{\sigma} E_0 d^{\dagger}_{\sigma}d_{\sigma}$.
The spin is fixed in the rotating frame and the exchange Hamiltonian is written as
$ 
\bar{H}_{SD} \mathord= \sum_{\sigma}\sigma v_s \mathrm{cos}(\vartheta)d^{\dagger}_{\sigma} d_{\sigma}\mathord+v_s \mathrm{sin}(\vartheta)  d^{\dagger}_{\uparrow}d_{\downarrow}\mathord+v_s \mathrm{sin}(\vartheta)d^{\dagger}_{\downarrow}d_{\uparrow}.
$
The remaining terms of the Hamiltonian \eqref{SQDSNMFH} are not affected by the transformation \eqref{UT}. Fig. \ref{LFRF} (b) depicts the system in the frame of the rotating spin. The splitting of the spin-up and spin-down energy levels of the quantum dot appears because of the exchange interaction and the energy shift of the spin-up (spin-down) quasiparticles is given by $ +(-)v_s\mathrm{cos}(\vartheta) $.

In order to evaluate the Green's functions of the system, we artificially divide the structure into three subsystems, which are the left lead (L), the right lead (R) and the quantum dot (D). \cite{Cuevas:2001km} 
The spin dependence and the superconducting state are taken into account by writing the Green's functions in Nambu-spin space. Since the system is out of equilibrium, we additionally write the Green's functions in Keldysh space. In Keldysh-Nambu-spin space, the Green's functions have the structure
\begin{equation}
\breve{G}_{\beta\beta^{\prime}}(t,t^{\prime}) = \begin{pmatrix} \hat{G}_{\beta\beta^{\prime}}^R & \hat{G}_{\beta\beta^{\prime}}^K \\ 0 & \hat{G}_{\beta\beta^{\prime}}^A \end{pmatrix}(t,t^{\prime}).
\label{KG}
\end{equation}
The symbols $ \, \breve{} \, $ and $ \, \hat{} \, $ denote a matrix in Keldysh-Nambu-spin and Nambu-spin space, respectively. The labels $ R $, $ A $ and $ K $ indicate the retarded, advanced and Keldysh element of the Green's function $ \breve{G}_{\beta\beta^{\prime}}(t,t^{\prime}) $. The indices $ \beta$ and $ \beta^{\prime} $ refer to operators in one of the three subsystems, e.g. the retarded Green's function $ \hat{G}^R_{LD} $ is given by $ \hat{G}^R_{LD}(t,t^{\prime}) \mathord= -i\theta(t-t^{\prime})\langle \{\psi_{L}(t),\psi^{\dagger}_{D}(t^{\prime}) \} \rangle $, where we have introduced the operators of the quantum dot and the leads in Nambu-spin space as $ \psi_{D}= ( d_{\uparrow} \,\,\, d_{\downarrow} \,\,\, d^{\dagger}_{\uparrow} \,\,\, d^{\dagger}_{\downarrow} )^T $ and $ \psi_{\alpha} =  ( c_{\alpha\uparrow} \,\,\, c_{\alpha\downarrow}  \,\,\, c^{\dagger}_{-\alpha\uparrow} \,\,\,  c^{\dagger}_{-\alpha\downarrow} )^T $ with the index $ \alpha $ referring to the momentum $ k_{\alpha} $ in the left or right lead. The elements of the matrix in \eqref{KG} are related by $ G^< =(1/2)\left( G^K-G^R+G^A \right)$ with the lesser Green's function defined by $ G_{DL}^<(t,t^{\prime}) = i \langle \psi_{L}^{\dagger}(t^{\prime})\psi_{D}(t)\rangle $. 

Taking into account all $ k $ states of the leads, we define the matrices $\textbf{G}_{\beta\beta^{\prime}} $ and $ \textbf{V}_{\beta\beta^{\prime}} $ as $ (\textbf{G}_{\beta\beta^{\prime}})_{ij} \mathord= \hat{G}_{{{\beta}_i},{\beta^{\prime}_j}} $ and   $ (\textbf{V}_{\beta\beta^{\prime}})_{ij} \mathord= \hat{V}_{{{\beta}_i},{\beta^{\prime}_j}}  $  with the diagonal matrices $ \hat{V}_{{\alpha},d} \mathord= \mathrm{diag}(  V_{d_{\uparrow},{\alpha}}, V_{d_{\downarrow},{\alpha}}, -V_{d_{\uparrow},-{\alpha}},   -V_{d_{\downarrow},-{\alpha}}) $ in Nambu-spin space. The indices $ i $ and $ j $ indicate all $ k $ states in the leads.  
The matrices \eqref{KG}  of all subsystems are then combined in an enlarged Hilbert space into one matrix defined by $ \tilde{{\textbf{G}}} $. The full and the unperturbed Green's function are written as\begin{displaymath}
\tilde{{\textbf{G}}}= \begin{pmatrix} \breve{\textbf{G}}_{LL} & \breve{\textbf{G}}_{LD}& \breve{\textbf{G}}_{LR} \\ \breve{\textbf{G}}_{DL}& \breve{\textbf{G}}_{DD} & \breve{\textbf{G}}_{DR}\\ \breve{\textbf{G}}_{RL} & \breve{\textbf{G}}_{RD}& \breve{\textbf{G}}_{RR}\end{pmatrix}, \tilde{{\textbf{G}}}_0= \begin{pmatrix} \breve{\textbf{G}}_{0L} & 0 & 0 \\ 0 & \breve{\textbf{G}}_{0D}& 0 \\ 0 & 0 & \breve{\textbf{G}}_{0R} \end{pmatrix}.
\end{displaymath} 
The coupling of the quantum dot to the leads and to the molecular magnet is given by 
\begin{displaymath} 
\tilde{{\textbf{V}}}= \begin{pmatrix} 0 & \breve{\textbf{V}}_{LD} & 0 \\ \breve{\textbf{V}}_{DL} & \breve{\textbf{V}}_{DD}& \breve{\textbf{V}}_{DR}\\ 0 & \breve{\textbf{V}}_{RD}& 0 \end{pmatrix},
\end{displaymath}
where $\hat{V}_{DD} =v_s\left(\hat{\sigma}_z\mathrm{cos}(\vartheta)+ \hat{\sigma}_x\mathrm{sin}(\vartheta)\right) $, is treated as a perturbation. The elements containing the coupling between the quantum dot and the leads are diagonal in Keldysh space and are in Nambu-space given by  $ \hat{V}_{{\alpha},d} $. 

Fourier transforming the Dyson equation to energy space, we obtain  \begin{equation}  \tilde{\textbf{G}}= \tilde{\textbf{G}}_0 + \tilde{\textbf{G}}_0 \tilde{\textbf{V}}\tilde{\textbf{G}} \label{Dyseqn}\end{equation} and can calculate the retarded and advanced Green's functions of the dot as
\begin{equation} \hat{G}^{R/A}_{DD} = \left(1-\hat{G}^{R/A}_{0D}\left(\hat{\Sigma}^{R/A}+\hat{V}_{DD}\right)\right)^{-1}\hat{G}^{R/A}_{0D},
\label{GDD}
\end{equation} 
where the self-energy is defined as $ \hat{\Sigma} = \hat{V}_{DL}\hat{g}_{LL}\hat{V}_{LD} + \hat{V}_{DR}\hat{g}_{RR}\hat{V}_{RD} =\hat{\Sigma}_L+\hat{\Sigma}_R$. These self-energies are obtained by summation over all quasiparticle states in the left and right leads, respectively. 
This summation can be replaced by the integral $ \sum_{k_{\alpha}} \rightarrow \frac{1}{(2\pi)^3} \int d^3k_{\alpha} \rightarrow N_0 \int d\xi_{\alpha} \int\frac{d\Omega_{k_{\alpha}}}{4\pi} $ with the normal density of states at the Fermi energy, $N_0 $.  The integration over the quasiparticle energies in the superconductor leads to the quasiclassical Green's function denoted as \cite{Kopnin:2009wt,Serene:1983vc}
\begin{displaymath}
\hat{g}_{\alpha}^{R/A} = \frac{-\pi}{\sqrt{\vert \Delta \vert^2-(E^{R/A})^2}}\begin{pmatrix} E^{R/A} &\Delta_{\alpha}i\sigma_y  \\ i\sigma_y\Delta^*_{\alpha}  & -E^{R/A} \end{pmatrix},
\end{displaymath}
together with the normalisation condition $ \left(\hat{g}_{\alpha}^{R/A}\right)^2 =-\pi^2\hat{1} $. 
The retarded and advanced self-energies are approximated by $  \hat{\Sigma}_{\alpha}^{R/A} =N_{\alpha} \vert \hat{V} \vert^2 \hat{g}_{\alpha}^{R/A}  $ and the energy $ E^{R/A} $ is defined as $ E^{R/A} = E\pm i\eta $ with $ \eta\to 0 $. In the rotating frame, the self-energies are given by (setting $ \tilde{\omega}=\omega_L/2$)
\begin{multline}
\hat{\Sigma}_{\alpha}^R = -\Gamma_{\alpha} \\ \begin{pmatrix} \frac{E^R+\tilde{\omega}}{\sqrt{\vert\Delta\vert^2\mathord-(E^R\mathord+\tilde{\omega})^2}} & \hspace{-0.7cm}0 &\hspace{-0.7cm} 0 & \hspace{-1cm}\frac{-\Delta_{\alpha}}{\sqrt{\vert\Delta\vert^2-(E^R+\tilde{\omega})^2}} \\ \hspace{-0.3cm} 0 & \hspace{-1cm} \frac{E^R-\tilde{\omega}}{\sqrt{\vert\Delta\vert^2-(E^R-\tilde{\omega})^2}} & \hspace{-0.2cm} \frac{\Delta_{\alpha}}{\sqrt{\vert\Delta\vert^2-(E^R-\tilde{\omega})^2}} & \hspace{-0.8cm} 0 \\ \hspace{-0.3cm} 0 &\hspace{-1cm} \frac{-\Delta^{\dagger}_{\alpha}}{\sqrt{\vert\Delta\vert^2-(E^R-\tilde{\omega})^2}} & \hspace{-0.2cm}\frac{-(E^R-\tilde{\omega})}{\sqrt{\vert\Delta\vert^2-(E^R-\tilde{\omega})^2}} & \hspace{-0.8cm}0 \\ \frac{\Delta^{\dagger}_{\alpha}}{\sqrt{\vert\Delta\vert^2-(E^R+\tilde{\omega})^2}} & \hspace{-0.7cm}0 & \hspace{-0.7cm}0 & \hspace{-1cm} \frac{-(E^R+\tilde{\omega})}{\sqrt{\vert\Delta\vert^2-(E^R+\tilde{\omega})^2}}  \end{pmatrix}\hspace{-0.1cm},
\label{selfenergy}
\end{multline}
with the tunnelling rates defined as $ \Gamma_{\alpha} = \pi N_0 V_{D\alpha}^2 $. 
The electron and hole part of the spin-up and spin-down unperturbed Green's function of the dot in Nambu-spin space is given by $ (\hat{G}_{0D}^{R})^{-1}_{11/22} \mathord=  E^R\mathord-E_0 $ and  $ (\hat{G}_{0D}^{R})^{-1}_{33/44}\mathord=  \mathord-(E^R+E_0) $. The unperturbed Green's function and the self-energy enable us to evaluate the full Green's function of the dot \eqref{GDD}. 

The average charge current operator in the Nambu-spin space from the left lead to the quantum dot is obtained by using the Heisenberg equation of motion 
\begin{equation}
\hat{J}_L = -i\frac{e}{\hbar} \left< \left[\hat{\bar{H}},\hat{\bar{N}}_L\right]\right> = -i \frac{e}{\hbar} \left< \left[\hat{\bar{H}}_T,\hat{\bar{N}}_L\right]\right>,
\label{HEOM}
\end{equation}
with the Hamilton operator $ \hat{\bar{H}} $ in the rotating frame, the number operator $ \hat{\bar{N}}_L = \frac{1}{2}\sum_{k_L}\psi_{k_L}^{\dagger}\hat{\sigma}_0\psi_{k_L}$ and the tunnelling operator $ \hat{\bar{H}}_{T\alpha} = \frac{1}{2}\sum_{k_{\alpha}} \psi_{k_{\alpha}}^{\dagger}\hat{V}_{d,k_{\alpha}}\psi_{d} + \psi^{\dagger}_{d}\hat{V}^{\dagger}_{d,k_{\alpha}} \psi_{k_{\alpha}} $.
The Josephson current \eqref{HEOM} can then be written in terms of the lesser Green's function as \cite{Jauho:1994te}
\begin{equation}
 J_L \mathord= \frac{e}{2\hbar}\int \frac{dE}{2\pi}\, \mathrm{Tr}\left(\hat{\sigma}_0\left(\textbf{G}_{DL}^<\textbf{V}_{LD}\mathord-\textbf{V}_{DL}\textbf{G}_{LD}^<\right)\right).
\label{KC}
\end{equation}
By using the Dyson equation \eqref{Dyseqn} in the enlarged Hilbert space, we calculate the elements $ \textbf{G}_{DL}^< $ and $ \textbf{G}_{LD}^< $ with the help of the relation $ G^< =(1/2)\left( G^K-G^R+R^A \right)$. The results are\cite{Cuevas:TshMQ34R}
\begin{eqnarray}
\textbf{G}_{DL}^<  &=& \textbf{G}_{DD}^<\textbf{V}_{DL}\textbf{g}_{LL}^A + \textbf{G}_{DD}^R\textbf{V}_{DL}\textbf{g}_{LL}^< \\ \textbf{G}_{LD}^< & = & \textbf{g}_{LL}^< \textbf{V}_{LD} \textbf{G}_{DD}^A + \textbf{g}_{LL}^R \textbf{V}_{LD} \textbf{G}_{DD}^<.
\end{eqnarray}
The Josephson current then simplifies to
\begin{equation}
 J_L \mathord= \frac{e}{\hbar}\int \frac{dE}{2\pi}\,\mathrm{Re}\left[\mathrm{Tr}\,\hat{\sigma}_0\left(\hat{G}_{DD}^R\hat{\Sigma}_L^< +\hat{G}_{DD}^< \hat{\Sigma}_L^A\right)\right].
 \label{CurGF}
\end{equation}
The Green's function $ \hat{G}_{DD}^< $ is obtained from the Keldysh equation by $ \hat{G}_{DD} ^<= \hat{G}_{DD}^R\hat{\Sigma}^<\hat{G}_{DD}^A $ with $ \hat{\Sigma}^< = \hat{\Sigma}^R \hat{F} \mathord- \hat{F} \hat{\Sigma}^A$. Due to the transformation to the rotating frame, the effective Fermi energy in Nambu-spin space are shifted and $ \hat{F} $ is given by 
\begin{equation} 
\hat{F} = \begin{pmatrix} f(E\mathord+\tilde{\omega})  & 0 & 0 & 0 \\ 0 & f(E\mathord-\tilde{\omega}) & 0 & 0 \\ 0 & 0 &   f(E\mathord-\tilde{\omega}) & 0 \\ 0 & 0 & 0 & f(E\mathord+\tilde{\omega}) \end{pmatrix}
\label{FeFu}
\end{equation}
with the Fermi function $ f(E) = 1/(1+\mathrm{exp}(E/k_BT)) $.

\section{Results}\label{sec:results}
The charge transport properties of the junction can now be investigated. In principle, the current is given by two contributions. The first is the current carried by Andreev states whose energies lie within the superconducting gap. The second contribution is carried by continuum states outside the superconducting gap.

\subsection{Density of states of the quantum dot} \label{subsec:DOS}
\begin{figure}[!b]
\begin{centering}
\subfigure{\hspace{-0.2cm}
\includegraphics[width=0.5\columnwidth]{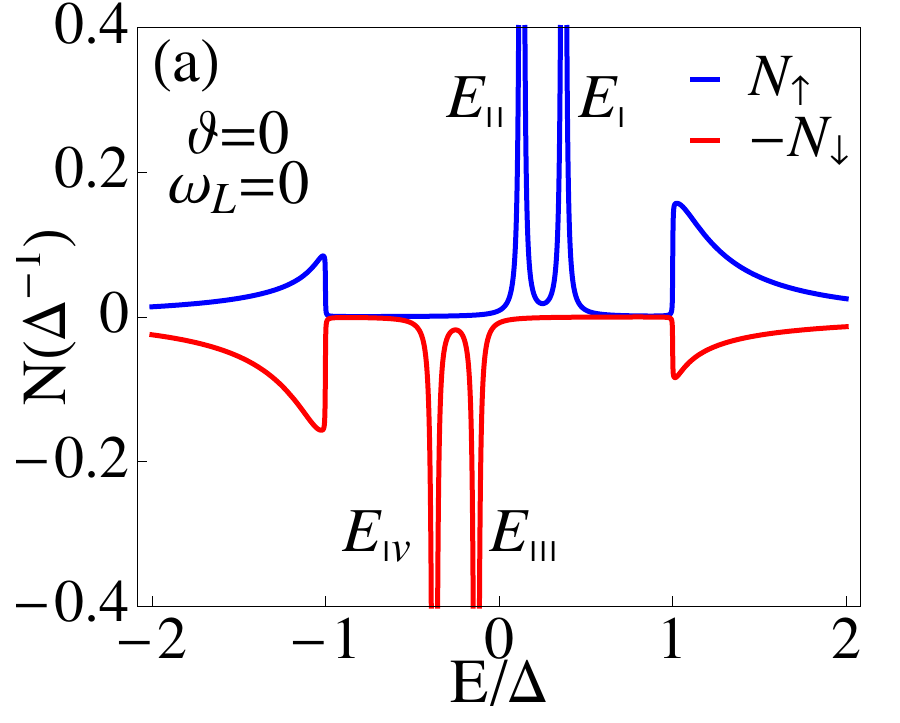}}\vspace{-0.2cm}
\subfigure{\hspace{-0.2cm}
\includegraphics[width=0.5\columnwidth]{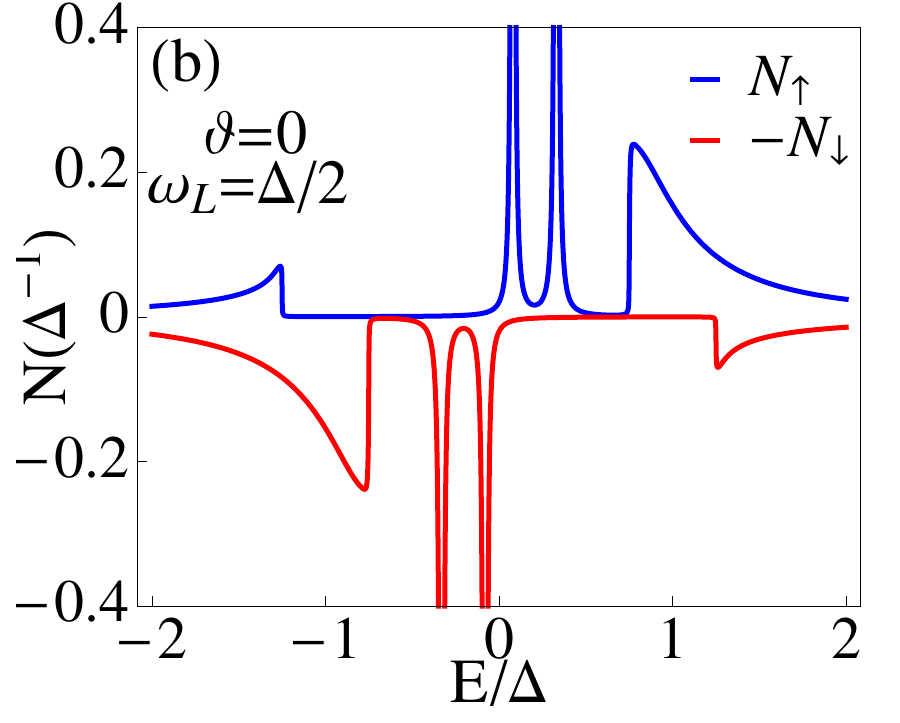}}\vspace{-0.0cm}
\subfigure{\hspace{-0.2cm}
\includegraphics[width=0.5\columnwidth]{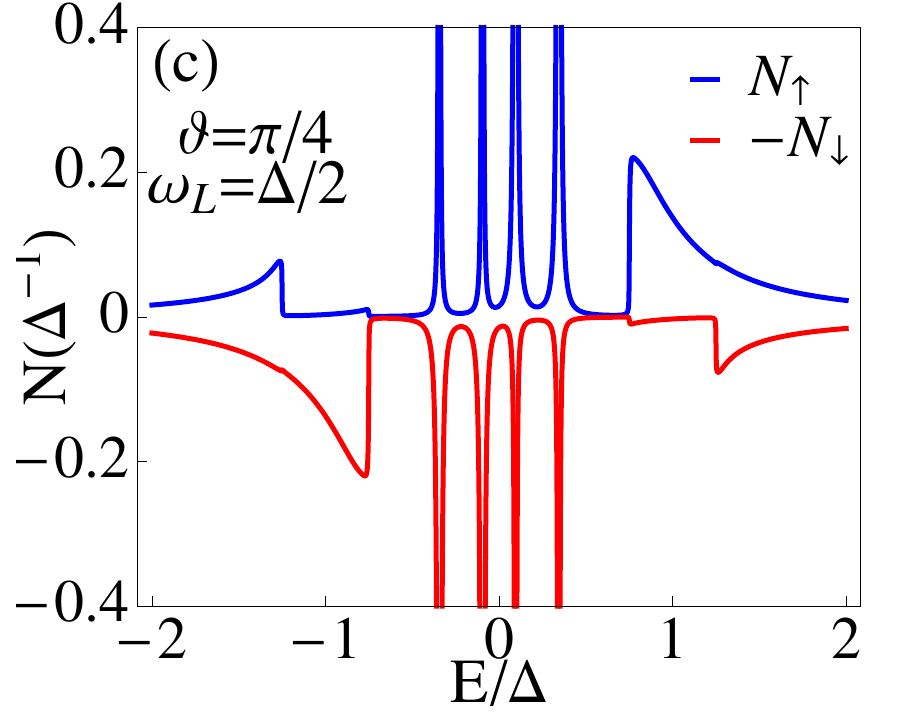}}\vspace{-0.2cm}
\subfigure{\hspace{-0.2cm}
\includegraphics[width=0.5\columnwidth]{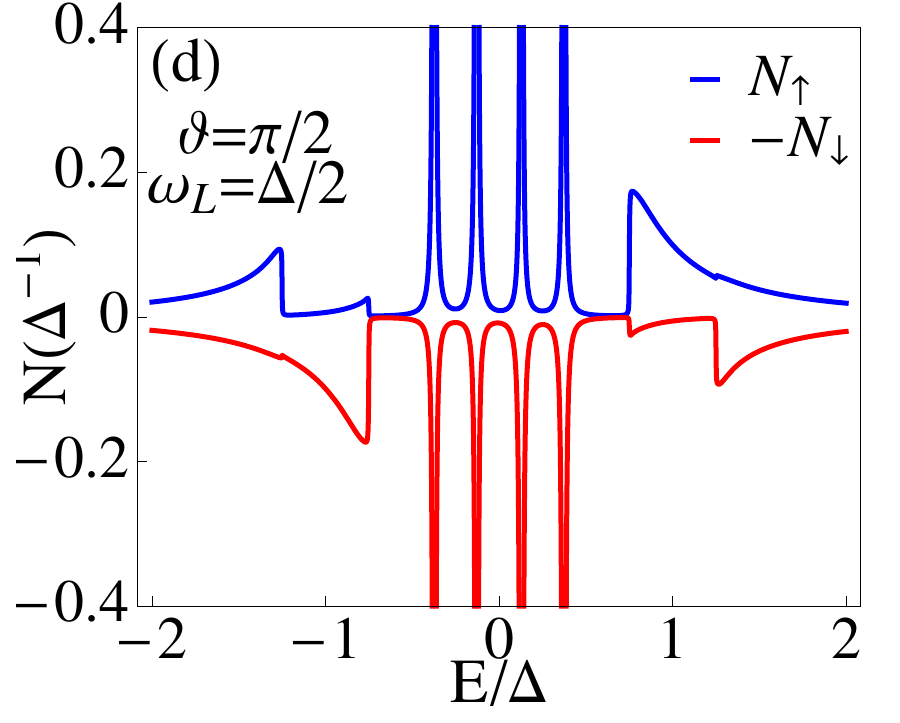}}\vspace{-0.0cm}
\caption{(Color online) Spin-resolved density of states of the quantum dot. In (a) and (b), $v_s=0.3\Delta$ and $\vartheta=0$. In (b) a magnetic field of $\omega_L=\Delta/2$ is applied, shifting the continuum states of the spin-up (spin-down) electrons by $\pm \omega_L/2$. In (c) and (d), $\vartheta\neq 0$ and an electron on the quantum dot can change its spin direction such that the states of the spin-up and spin-down electrons are mixed. In (d), the magnetisation is precessing in the plane ($\vartheta=\pi/2$). The other parameters in panel (a)-(d) are $\varphi=\pi/2$, $\Gamma=0.1\Delta$, $E_0=0$ and $\eta=10^{-3}\Delta$.}
\label{figDensity}
\end{centering}
\end{figure}

Figure \ref{figDensity} shows the spin-resolved density of states of the quantum dot at  $ \varphi=\pi/2 $, $ E_0=0 $ and a symmetric coupling to the leads with  $ \Gamma=\Gamma_L=\Gamma_R=0.1\Delta $. In panel (a), we consider a static magnetisation of the molecular magnet in the $ z $ direction ($ \omega_L=0, \vartheta=0 $). The exchange interaction between the molecular magnet and the quantum dot lifts the spin degeneracy and shifts the energy level of the quantum dot in the rotating frame by $ \pm v_s $ for spin-up and spin-down electrons, respectively. Therefore, the spin-up energy level, which is broadened by $ \Gamma $, is closer to the upper superconducting gap edge increasing the density of the spin-up continuum states for $ E > \Delta $ whereas the density of the spin-down continuum states is decreased. The exchange interaction also lifts the spin degeneracy of the Andreev states and shifts the states $ E_{\Romannum{1}} $ and $ E_{\Romannum{2} }$ ($ E_{\Romannum{3}} $ and $ E_{\Romannum{4} }$ ) of the spin-up (spin-down) quasiparticles to higher (lower) energies. For the parameters in panel (a), the exchange coupling pushes both spin-up (spin-down) states above (below) $ E = 0 $. 

In panel (b), a magnetic field is applied in addition to the exchange coupling of the molecular magnet and the quantum dot. The magnetic field shifts the continuum states by $ -(+) \omega_L/2 $ for spin-up (spin-down) electrons due to the transformation in the rotating frame and also slightly pushes the Andreev states towards $ \vert E \vert \rightarrow 0 $ compared to panel (a). If the magnetisation points in an arbitrary direction ($ \vartheta\neq 0 $), the electrons can undergo spin flips into sidebands separated by the energy   $ \hbar \omega_L $ according to the Hamiltonian \eqref{SQDSNMFH}. In the frame of the rotating spin, the transformation compensates the exchange of energy and the electrons are scattered into states at the same energy. In panel (c), $ \vartheta=\pi/4 $  and the Andreev states as well as the continuum states can be occupied with spin-up and spin-down electrons respectively. In panel (d), $ \vartheta=\pi/2 $, the magnetisation is precessing in the $ xy $ plane and the density of the scattered states increases. 

The complete parameter dependence of the Andreev states in the central region are obtained by the poles of the Green's function in Eq. \eqref{GDD} which are given by
\begin{equation} 
\mathcal{A}_+\mathcal{A}_--\mathcal{B}=0.
\label{AS} 
\end{equation}
Using the notation $ \tilde{\omega} = \omega_L/2 $, the elements are 
\begin{multline*}
\mathcal{A}_{\pm} = \left((E\mp v_s\mathrm{cos}\vartheta )\sqrt{\Delta^2-(E\pm \tilde{\omega} )^2}+2\Gamma(E\pm\tilde{\omega} )\right)^2\\-4\Gamma^2\Delta^2\mathrm{cos}^2(\varphi/2) -E_0^2\left(\Delta^2-(E\pm\tilde{\omega})^2\right),
\end{multline*}
and
\begin{multline*}
\mathcal{B} =v_s^2\mathrm{sin}^2\vartheta \sqrt{(\Delta^2-(E-  \tilde{\omega})^2)(\Delta^2-(E+ \tilde{\omega})^2)} \\ \bigg[8\Gamma^2\left(\Delta^2\mathrm{cos}^2(\varphi/2)+\left(E^2-\tilde{\omega}^2\right)\right) \\ +\sqrt{\Delta^2-(E-\tilde{\omega})^2}\sqrt{\Delta^2-(E+ \tilde{\omega} )^2}\left(-v_s^2\mathrm{sin}^2\vartheta \right. \\ \left.+2\left(E^2+E_0^2-v_s^2\mathrm{cos}^2\vartheta\right)\right)+\\4\Gamma\left((E+ \tilde{\omega} )(E+v_s\mathrm{cos}\vartheta)\sqrt{\Delta^2-(E-\tilde{\omega} )^2}\right. \\[-0.2cm] \left.+(E- \tilde{\omega})(E-v_s\mathrm{cos}\vartheta)\sqrt{\Delta^2-(E+\tilde{\omega} )^2}\right)\bigg].
\end{multline*}
The pole equation \eqref{AS} reduces to that of Ref. [\onlinecite{Beenakker:2001wo}], if the quantum dot does not interact with the molecular magnet and no magnetic field is applied ($ v_s=\tilde{\omega}=0 $). In this case $\mathcal{A}_+$ is equal to $\mathcal{A}_-$ and $\mathcal{B} = 0$. If no magnetic field is applied, the $ \vartheta $-dependence of the Andreev states vanishes since no spin-quantisation axis is preferred. In this limit, the equation of the Andreev states agrees with the result in Ref. [\onlinecite{Benjamin:2007fz}]. If $ \vartheta=\pi/2 $, the spin precesses in the $ xy $ plane and the equation of the Andreev states is symmetric with respect to $ \omega_L \rightarrow -\omega_L $. Additionally, the Andreev states are symmetric under the transformation $ E_0 \rightarrow -E_0 $.  

\begin{figure}[!t]
\begin{centering}
\subfigure{\hspace{-0.2cm}
\includegraphics[width=0.5\columnwidth]{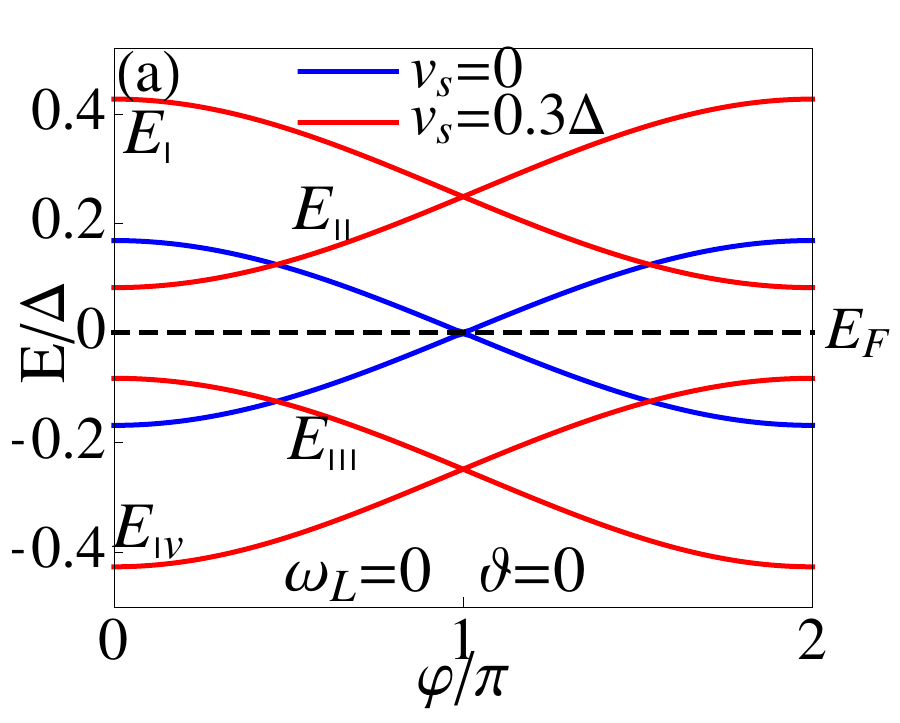}}\vspace{-0.2cm}
\subfigure{\hspace{-0.2cm}
\includegraphics[width=0.5\columnwidth]{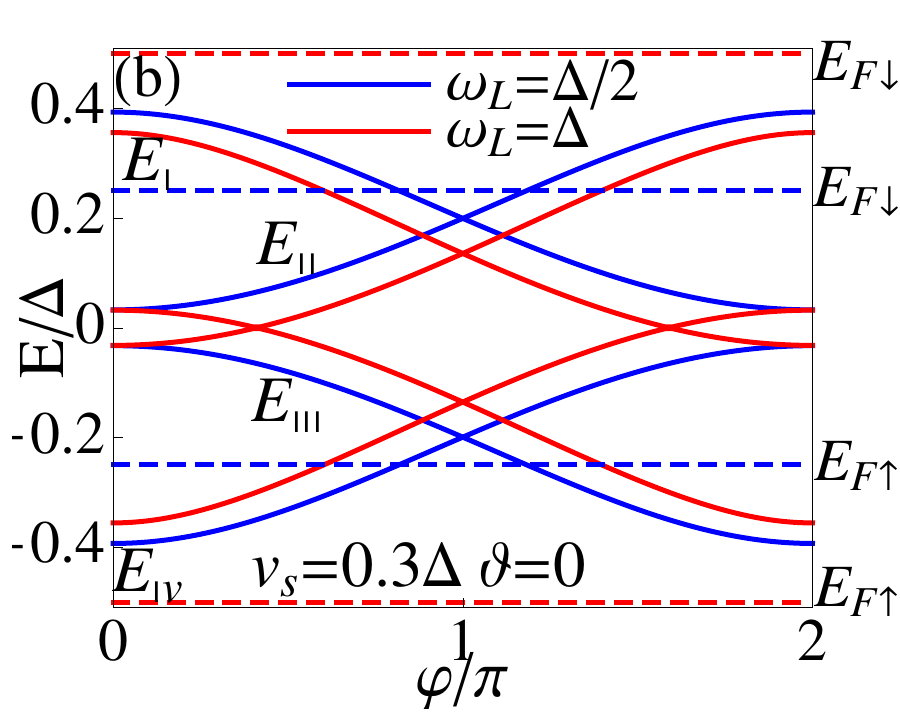}}\vspace{-0.0cm}
\subfigure{\hspace{-0.2cm}
\includegraphics[width=0.5\columnwidth]{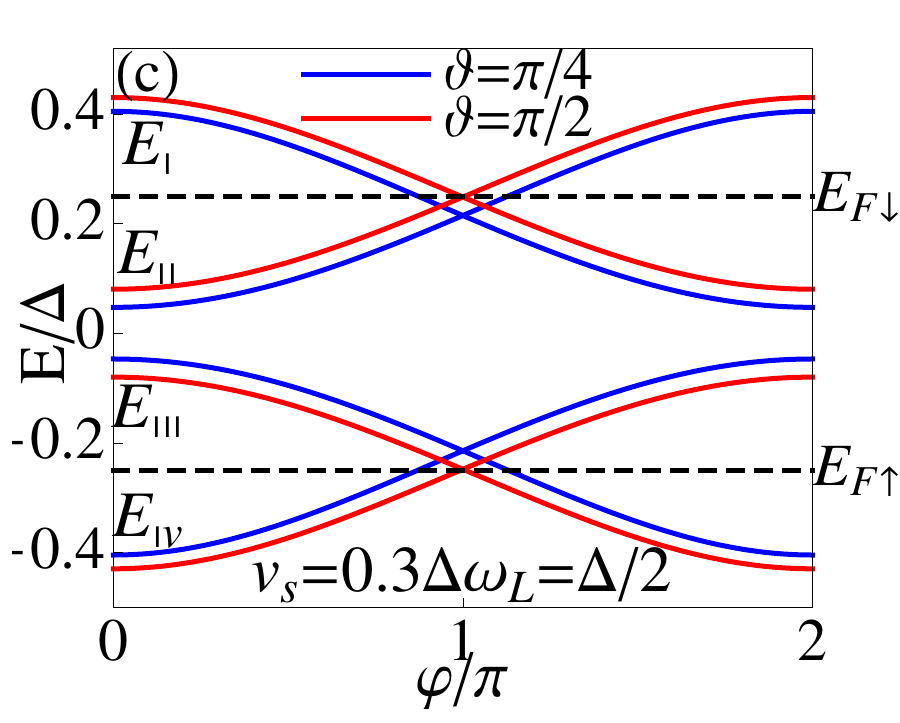}}\vspace{-0.2cm}
\subfigure{\hspace{-0.2cm}
\includegraphics[width=0.5\columnwidth]{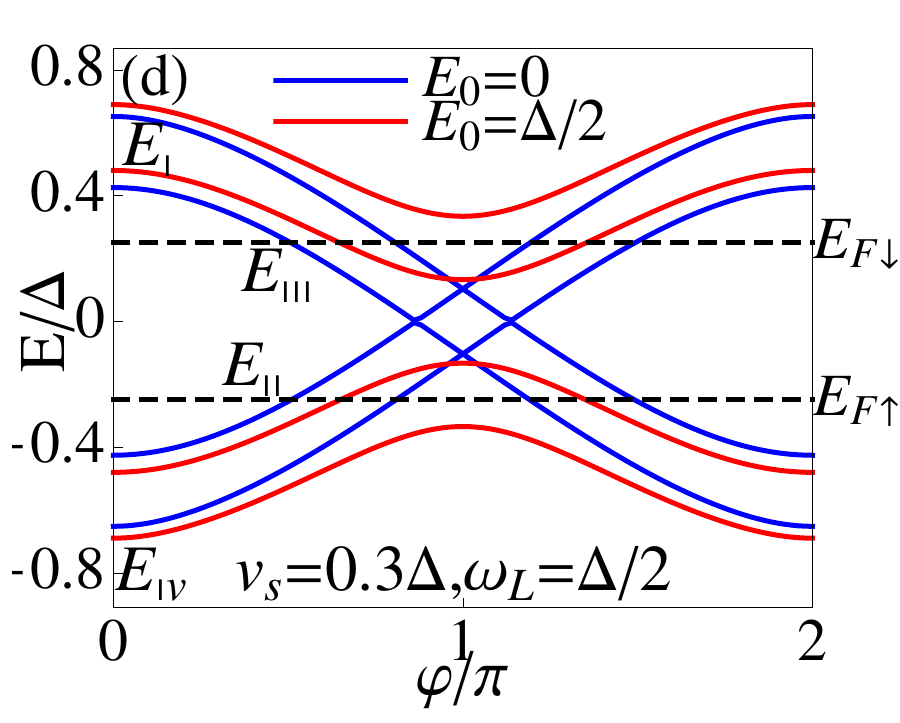}}\vspace{-0.0cm}
\caption{(Color online) Phase dependence of the Andreev states. In panels (a)-(c), $\Gamma=\Delta/10$ and $E_0 = 0$. In (a), the exchange coupling $v_s$ splits the Andreev state into spin-up and spin-down states. The effect of a finite $\vartheta$ is shown in panel (c). Panel (d) shows the Andreev states for $\Gamma=\Delta/2$, $\vartheta=\pi/4$ and a position of the energy level on the dot of $E_0=0$ and $E_0=\Delta/2$. The dashed lines indicate the effective Fermi energies of the spin-up and spin-down quasiparticles at zero temperature.}
\label{figABS}
\end{centering}
\end{figure}

In the limit  $ E \ll \Delta $, the equation of the Andreev states can be explicitly solved with the result
\begin{multline}
E_{\Romannum{1},\Romannum{2}} \approx \frac{1}{v}\left(\pm \sqrt{\frac{\left(2 \Gamma \Delta \mathrm{cos}(\varphi/2)\right)^2 }{\Delta^2\mathrm-\tilde{\omega}^2}\mathord+E_0^2} \right. \\ \left. \mathord+\sqrt{\left(v_s \mathrm{cos}(\vartheta)\mathord-\frac{\Gamma \omega}{\sqrt{\Delta^2\mathord-\tilde{\omega}^2}}\right)^2+ v_s^2 \mathrm{sin}^2(\vartheta)}\right)
\label{absappa}
\end{multline}
and
\begin{multline}
E_{\Romannum{3},\Romannum{4}} \approx \frac{1}{v}\left(\pm \sqrt{\frac{\left(2 \Gamma \Delta \mathrm{cos}(\varphi/2)\right)^2 }{\Delta^2\mathrm-\tilde{\omega}^2}\mathrm+E_0^2} \right. \\ \left. \mathrm-\sqrt{\left(v_s \mathrm{cos}(\vartheta)\mathord-\frac{\Gamma \omega}{\sqrt{\Delta^2\mathrm-\tilde{\omega}^2}}\right)^2+ v_s^2 \mathrm{sin}^2(\vartheta)}\right)
\label{absappb}
\end{multline}
\noindent with $v=\sqrt{\vert \Delta \vert^2-\tilde{\omega}^2}/\left(\sqrt{\vert \Delta \vert^2-\tilde{\omega}^2}+2\Gamma\right) $.
In general, however, the Andreev states must be calculated numerically from Eq. \eqref {AS} and are shown in Fig. \ref{figABS}.
In panel (a), the spin degeneracy is lifted due to the exchange interaction between the quantum dot and the molecular magnet. The spin-up (spin-down) electrons are shifted to higher (lower) energies. Since the Andreev states lie well inside the energy gap, we can use Eqs. \eqref{absappa} and \eqref{absappb} to find expressions for the splitting of the Andreev states. The displacement of the Andreev states in panel (a) due to the exchange interaction is given by $ \pm v_s/(\Delta+2\Gamma) $. The exchange interaction shifts the Andreev states across the Fermi energy, which from equation \eqref{FeFu} is located at $ E=0 $ at $ T=0 $. Therefore, the current is expected to be strongly modified in panel (a) if the coupling $ v_s $ is increased. An applied magnetic field counteracts the shift of the Andreev states induced by the exchange interaction, $ v_s $. The combined shift of the Andreev states due to $ v_s $ and $ \omega_L $ is given by $ \pm(1/v)(v_s -\Gamma \tilde{\omega}/(\sqrt{\Delta^2\mathrm-\tilde{\omega}^2})) $. The effective Fermi energies in panel (b) at  $ T = 0 $ and $\omega_L=\Delta/2$ are located at the energies $ -(+) \omega_L/2 $ for spin-up (spin-down) electrons. In this case, both spin-up Andreev states are shifted above the effective Fermi energy $ E_{F\uparrow} $, whereas both spin-down Andreev states are below the effective Fermi energy $ E_{F\downarrow} $. The effect of $ \vartheta $ on the Andreev states is shown in panel (c). Now, all Andreev states below $ E_{F\downarrow} $ ($ E_{F\uparrow} $) are occupied with spin-down (up) electrons similarly to the situation of the density of states in Fig. \ref{figDensity} (c) and (d). In (d), $ \Gamma = \Delta/2 $ and the shift of the Andreev states due to $ \Gamma $ is larger than the splitting due to $ v_s $ such that at $ \varphi=0 $ the states $ E_{\Romannum{2}} $ and $ E_{\Romannum{4}} $ are below the effective Fermi energy $ E_{F\downarrow} $ and the states  $ E_{\Romannum{1}} $ and $ E_{\Romannum{3}} $ are above the effective Fermi energy $ E_{F\uparrow} $.  

\subsection{Current-phase relations}\label{subsec: current phase relation}
\begin{figure}[!t]
\begin{centering}
\subfigure{\hspace{-0.2cm}
\includegraphics[width=0.5\columnwidth]{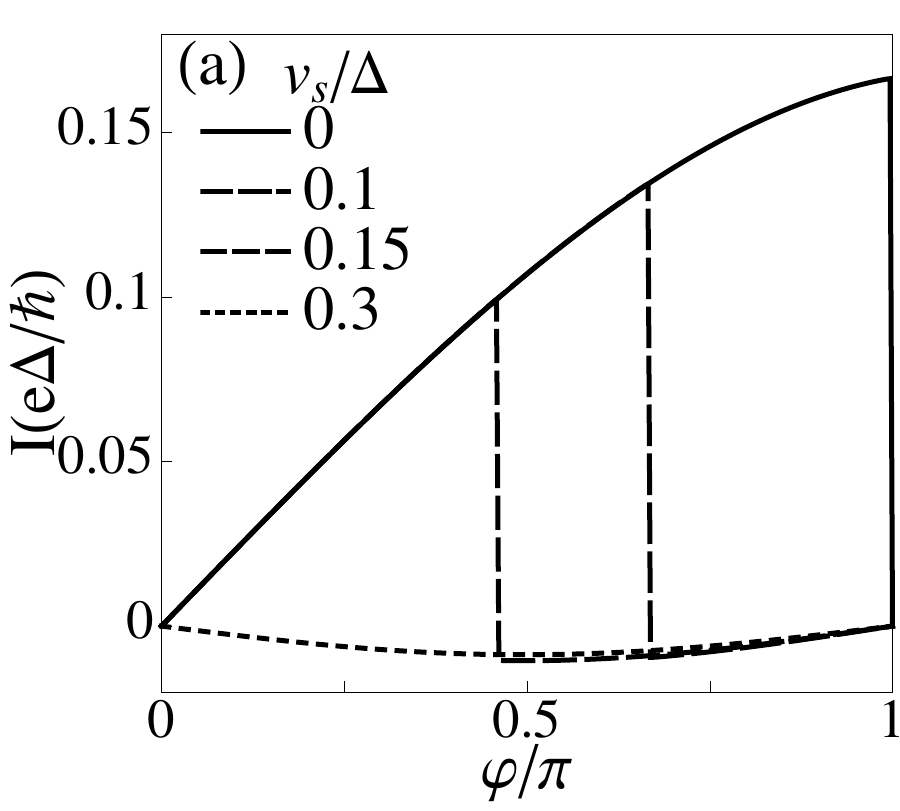}}\vspace{-0.2cm}
\subfigure{\hspace{-0.2cm}
\includegraphics[width=0.5\columnwidth]{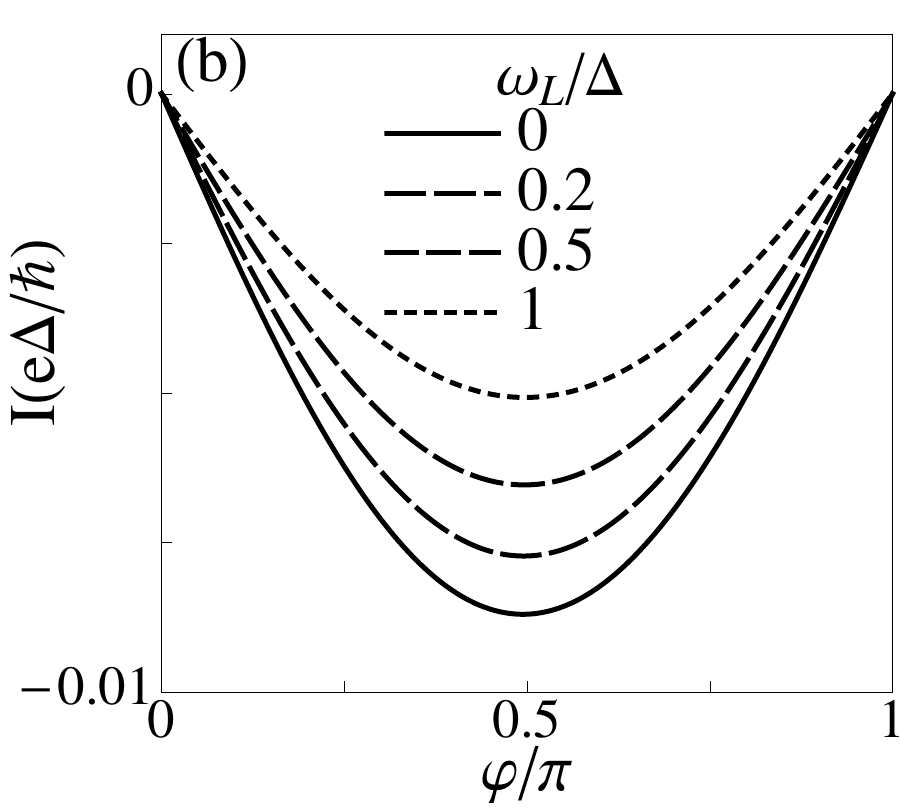}}\vspace{-0.1cm}
\subfigure{\hspace{-0.2cm}
\includegraphics[width=0.5\columnwidth]{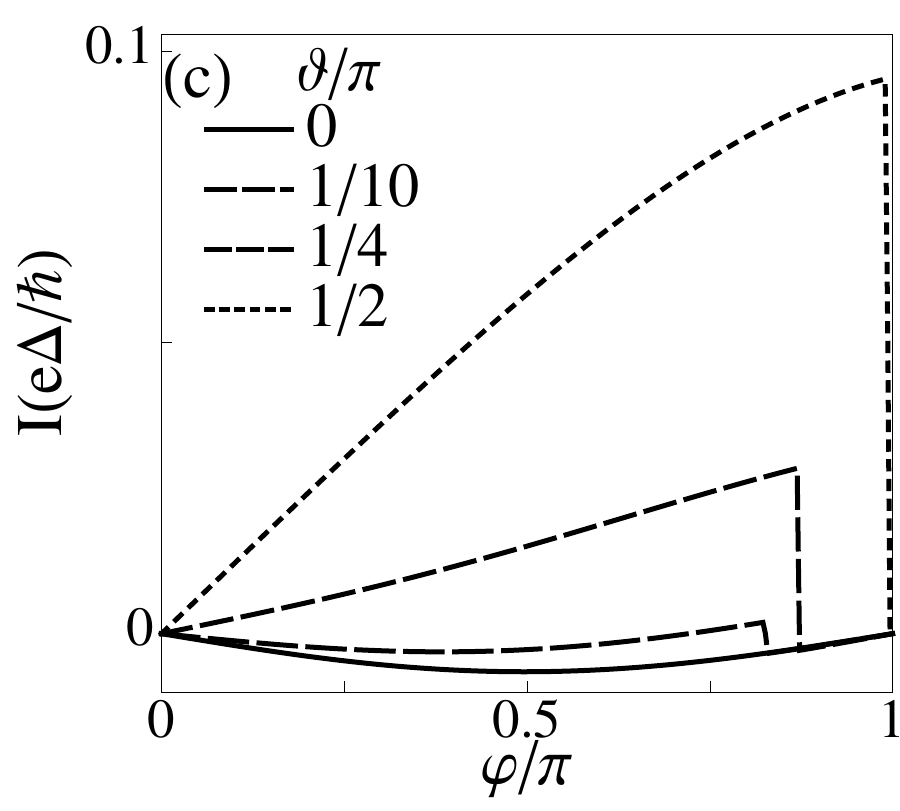}}
\subfigure{\hspace{-0.2cm}
\includegraphics[width=0.5\columnwidth]{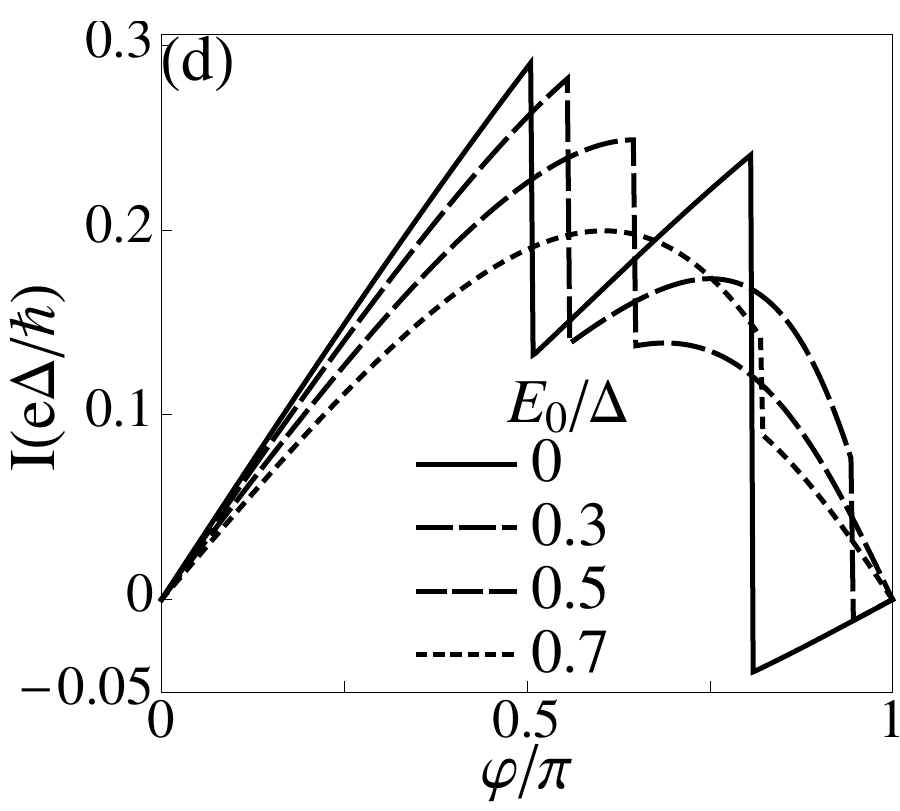}}
\caption{Current-phase relations for $\Gamma=\Delta/10$ and $E_0=0$ (panel (a)-(c)). In (a) $\omega_L=0$ and $\vartheta=0$ and $v_s$ varies from $0$ to $0.3\Delta$. The values correspond to the Andreev states of Fig. \ref{figABS} (a), where the current is strongly suppressed due to the shift of the Andreev states across the Fermi energy. In panel (b), $v_s=0.3\Delta$, $\vartheta=0$ and $\omega_L$ varies from $0$ to $\Delta$. In (c), $v_s=0.3\Delta$, $\omega_L=\Delta/2$ and $\vartheta$ varies from $0$ to $\pi/2$. In (d), $v_s=0.3\Delta$, $\omega_L=\Delta/2$, $\vartheta=\pi/4$, $\Gamma=\Delta/2$ and $E_0$ is varied. The temperature is set to $k_BT = 10^{-4}\Delta$ and $\eta=10^{-4}\Delta$.}
\label{figCPRABS}
\end{centering}
\end{figure}

\begin{figure*}[t]
\begin{centering}
\subfigure{\hspace{-0.2cm}
\includegraphics[width=1\columnwidth]{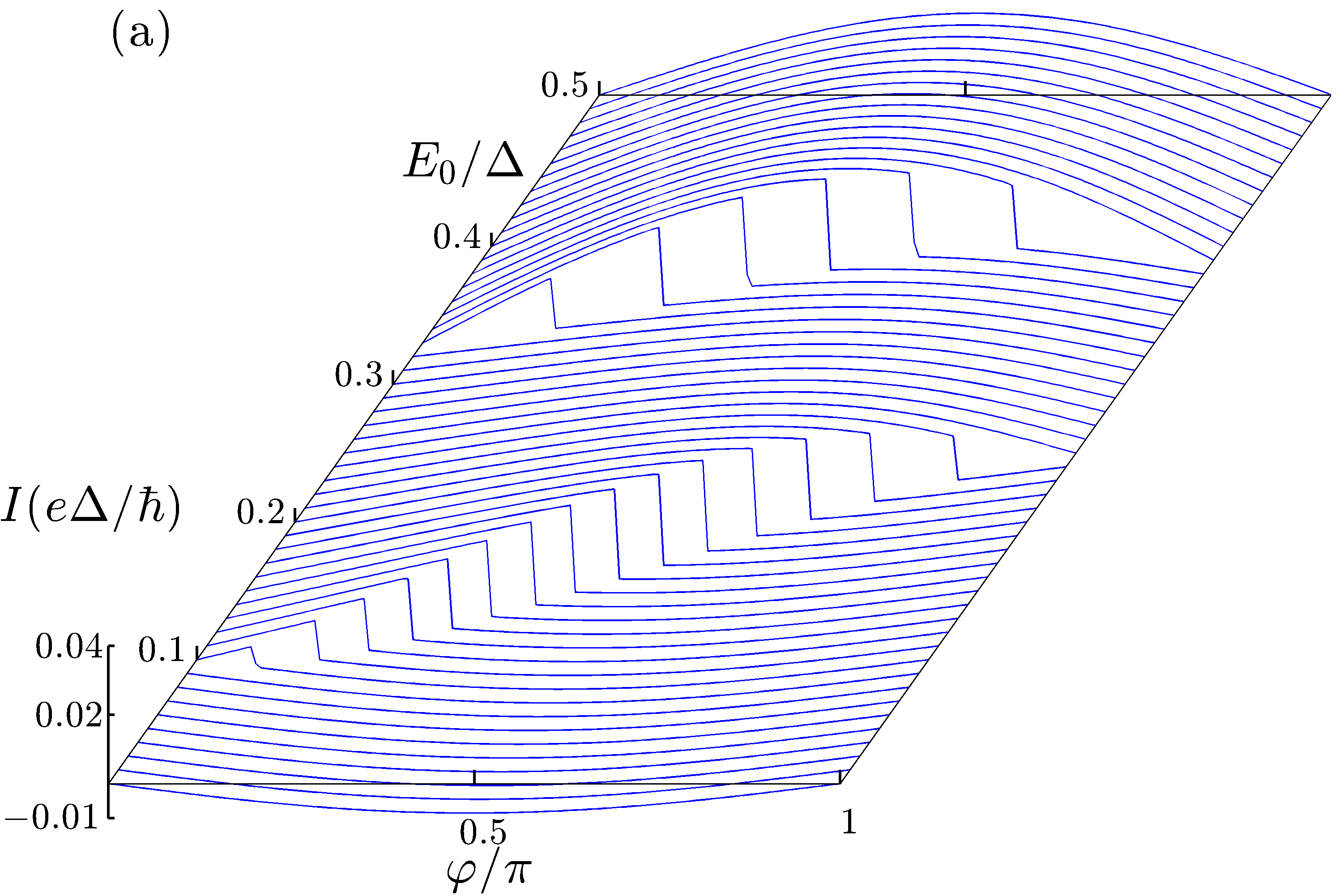}}\vspace{-0.2cm}
\subfigure{\hspace{-0.2cm}
\includegraphics[width=1\columnwidth]{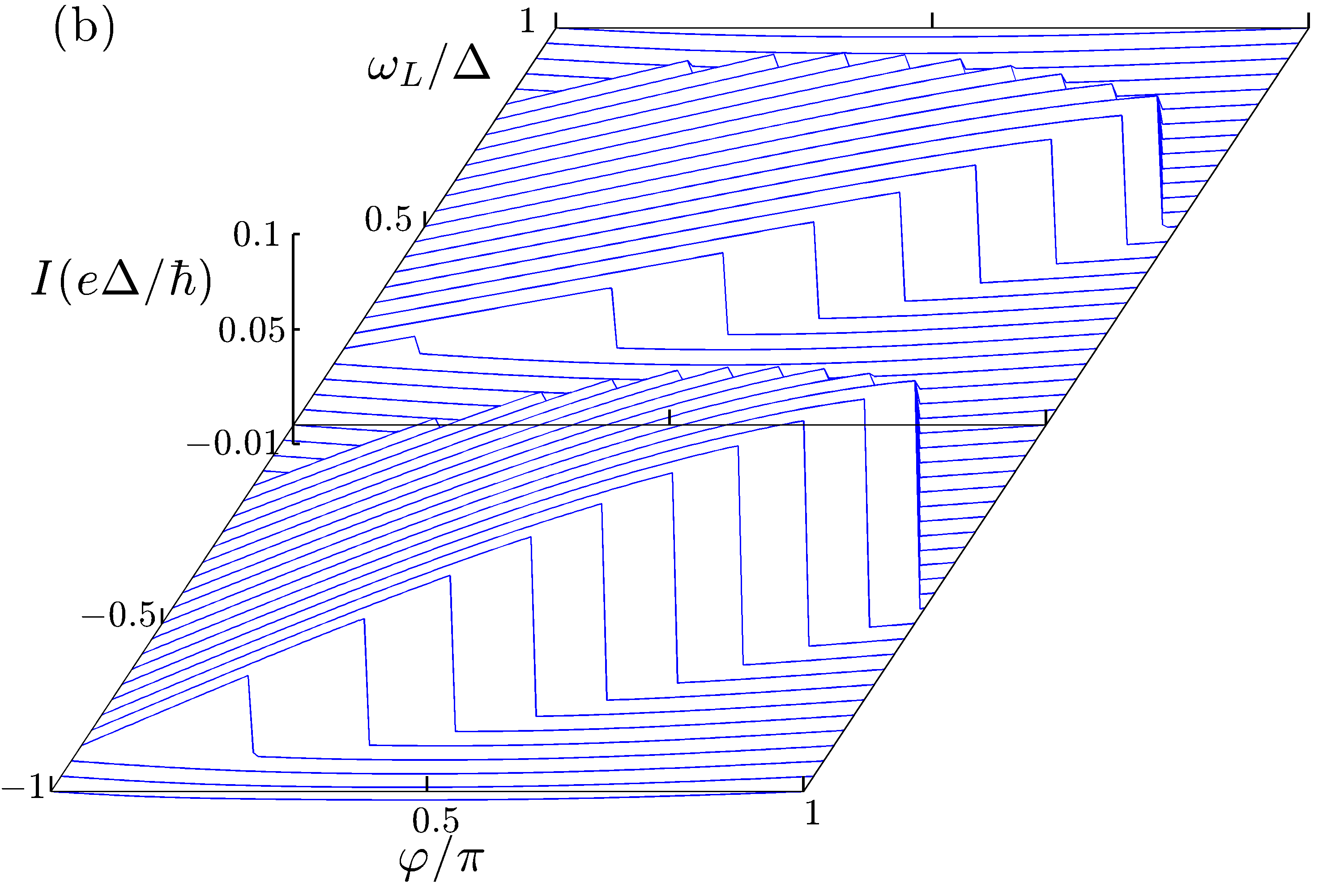}}\vspace{-0.1cm}
\caption{Current-phase relation for $v_s=0.1\Delta$, $\vartheta=\pi/4$, $\omega_L=\Delta/2$ (a). By changing $E_0$, the state changes from a $0$ to a $\pi$ state. In (b), $E_0=0$,$v_s=0.3\Delta$ and $\vartheta=0.4\pi$. The magnetic field changes from $\omega_L=-\Delta$ to $\omega_L=\Delta$. In both panels $\Gamma = \Gamma_L=\Gamma_R=0.1\Delta$, $k_BT = 10^{-4}\Delta$ and $\eta=10^{-4}\Delta$.}
\label{figCPRvb}
\end{centering}
\end{figure*}

The Andreev states discussed in the last section carry the Josephson current $ I_n(\varphi) $.
The contribution to the current of each Andreev state $ E_n(\varphi) $ is proportional to the derivative of the Andreev state with respect to the phase multiplied with the Fermi function at the energy of the Andreev state, \cite{Golubov:2004zz}
\begin{equation}
I_n(\varphi) = \frac{e}{\hbar} f(E_n(\varphi)) \frac{dE_n(\varphi)}{d\varphi}.
\label{oneabs}
\end{equation}
The Josephson current is then given by the summation over all Andreev states $ n $.

In the following, we discuss the current-phase relation obtained by Eq. \eqref{CurGF} in the low temperature limit $ k_BT = 10^{-4}\Delta $. This expression contains the contribution to the current from Andreev as well as the continuum states. Fig. \ref{figCPRABS} (a) shows the current-phase relation of a static spin in the $ z $ direction ($ \vartheta=0 $) and zero magnetic field ($ \omega_L=0 $) and four different values of  $ v_s $. The Andreev states corresponding to the values of $ v_s=0 $ and $ v_s=0.3\Delta $ are depicted in Fig. \ref{figABS} (a). At $  v_s=0 $, the spin-degenerate Andreev state below the Fermi energy $ E_F $ carries a positive current for $ \varphi < \pi $ according to relation \eqref{oneabs}. If the exchange coupling is increased to $ v_s=0.3\Delta $, the contributions from the current of both current-carrying Andreev states  $ E_{\Romannum{3}} $ and $ E_{\Romannum{4}} $ in Fig. \ref{figABS} (a) cancel whereas the Andreev states $ E_{\Romannum{1}} $ and $ E_{\Romannum{2}} $ are completely unoccupied above the Fermi energy. In this case, the current is strongly suppressed since in total the Andreev states do not contribute to the current and the current is carried by the continuum states giving rise to a $ \pi $ state of the junction. Between $ v_s=0 $ and the complete shift of the $ E_{\Romannum{3}} $ state below the Fermi energy, the current-phase relation sharply decreases at phases where the $ E_{\Romannum{3}} $ level intersects the Fermi energy. These phases are determined from Eq. \eqref{AS} to $ \varphi = 2\mathrm{arccos}(v_s/2\Gamma) $. At phases larger than $ 2\mathrm{arccos}(v_s/2\Gamma)  $, a negative current appears due to the continuum states since the Andreev-state contributions to the current cancel.

Figure \ref{figCPRABS} (b) shows the current-phase relation for the same parameters as in panel (a) at $ v_s=0.3\Delta $ and $ \vartheta=0 $, but the magnetic field is increased from $ \omega_L=0 $ to $ \omega_L=\Delta $. The Andreev states corresponding to the current-phase relation at $ \omega_L=\Delta/2 $ and $ \omega_L=\Delta $ are depicted in Fig. \ref{figABS} (b). For the parameters chosen in \ref{figCPRABS} (b), the spin-up (spin-down) Andreev states do not cross the corresponding effective Fermi energy $ E_{F\uparrow} $ ($ E_{F\downarrow} $) and no sharp change of the current is observed. 
The Andreev states $ E_{\Romannum{1}} $ and $ E_{\Romannum{2}} $ [Fig. \ref{figABS} (b)], which carry spin-up electrons, are for all values of $ \omega_L $ in Fig. \ref{figCPRABS} (b) above the effective Fermi energy $ E_{F\uparrow} $ and therefore do not contribute to the current. The Andreev states $ E_{\Romannum{3}} $ and $ E_{\Romannum{4}} $ carrying spin-down electrons are both below the effective Fermi energy $ E_{F\downarrow} $ and do not contribute either due to current cancelation of  $ E_{\Romannum{3}} $ and $ E_{\Romannum{4}} $ state. Therefore, only the continuum states give rise to a current such that the junction is in the $ \pi $ state.

In Fig. \ref{figCPRABS} (c), the current-phase relation is shown for the same parameters as in panel (b) at $ \omega_L=\Delta/2 $ but the angle $ \vartheta $ increases from $ 0 $ to $ \pi/2 $. Since the electrons on the quantum dot can undergo spin flips at finite $\vartheta $, the Andreev states in the rotating frame are degenerate and we have to take into account four spin-degenerate Andreev states following from the density of states in Fig. \ref{figDensity} (c) and (d). The Andreev states corresponding to the current-phase relation in Fig. \ref{figCPRABS} (c) at $ \vartheta=\pi/4 $ and $ \vartheta=\pi/2 $ are shown in Fig. \ref{figABS} (c). At $ \vartheta = 0 $, the current is the same as in panel (b) at $ \omega_L=\Delta/2 $ where $ E_{\Romannum{1}} $ and $ E_{\Romannum{2}} $  are empty while  $ E_{\Romannum{3}} $ and $ E_{\Romannum{4}} $ are occupied with spin-down electrons and therefore the Andreev states do not contribute to the current. If $ \vartheta $ is increased, spin-down (spin-up) electrons are scattered into $ E_{\Romannum{1}} $ and $ E_{\Romannum{2}} $ ($ E_{\Romannum{3}} $ and $ E_{\Romannum{4}} $) states. The particles scattered into the $ E_{\Romannum{2}} $ and $ E_{\Romannum{4}} $  states are below the corresponding effective Fermi energy $ E_{F\downarrow} $ and $ E_{F\uparrow} $, respectively. These two states give a positive contribution and therefore a finite $ \vartheta $ increases the current. The sharp step for phases close to $ \pi $ of the current-phase relation appears because an Andreev state crosses the Fermi energy, similarly as in panel (a). For $ \Gamma=\Delta/2 $, the current-carrying Andreev states cross the Fermi energy twice thus leading to the two steps in the current-phase relation as shown in panel (d). A finite $ E_0 $ opens a gap of the Andreev states at $ \varphi=\pi $. 

So far we have discussed the current-phase relations of the Andreev states shown in Fig. \ref{figABS}. We now consider how the $ 0 $ to $ \pi $ transition and the reverse process are driven by $ E_0 $ or $ \omega_L $. The current-phase relation as a function of $ E_0 $ is shown in Fig. \ref{figCPRvb} (a). The state changes from a $ \pi $ junction at $ E_0=0$ to a $ 0 $ junction at $ E_0=\Delta $. The current increases stepwise as a function of $ \varphiÊ$, but the position of this stepwise increase changes as a function of $ E_0 $. 
This behaviour appears twice, since the spin degeneracy of the Andreev states is lifted due to the exchange interaction. Since the current is symmetric under the transformation $ E_0 \rightarrow -E_0 $, the same transition is driven by decreasing $ E_0 $ from $ E_0=0 $ to $ E_0=-\Delta $. In panel (b) of Fig. \ref{figCPRvb}, the current-phase relation is evaluated as a function of $ \omega_L $. If the magnetic field is increased from $ \omega_L = -\Delta $ to $ \omega_L=0 $, the junction is driven from a $ \pi $ to $ 0 $ and back to a  $ \pi $ state. Since the Andreev states are symmetric under the transformation $ \omega_L \rightarrow -\omega_L $ if $ \vartheta = 0.5 \pi $ (Sec. \ref{subsec:DOS}), the Josephson current shows a similar behaviour if $ \vartheta $ approaches $ 0.5 \pi $.

\subsection{Critical Current}\label{subsec:spin currents and spin-triplet correlations}
As we discussed in the previous section, different parameters can drive the junction from a $ 0 $ to a $\pi$ state or vice versa. In order to further investigate the transport properties, we consider the critical current, which is experimentally more easily accessible than the current-phase relation. 

\begin{figure}[!b]
\begin{centering}
\subfigure{\hspace{-0.2cm}
\includegraphics[width=0.5\columnwidth]{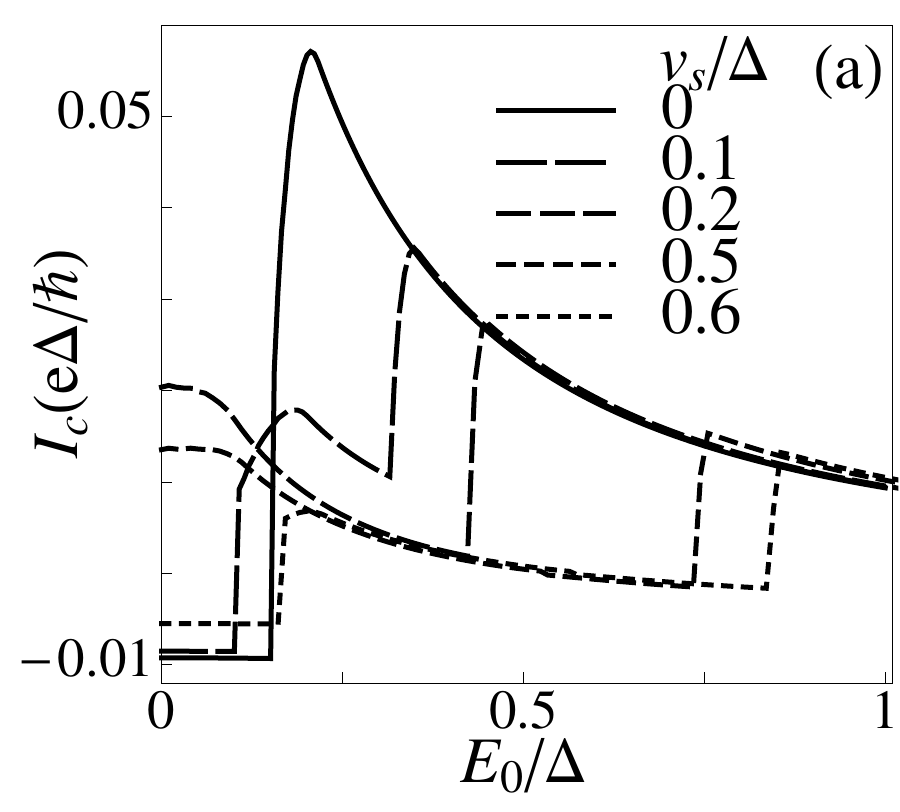}}\vspace{-0.2cm}
\subfigure{\hspace{-0.2cm}
\includegraphics[width=0.5\columnwidth]{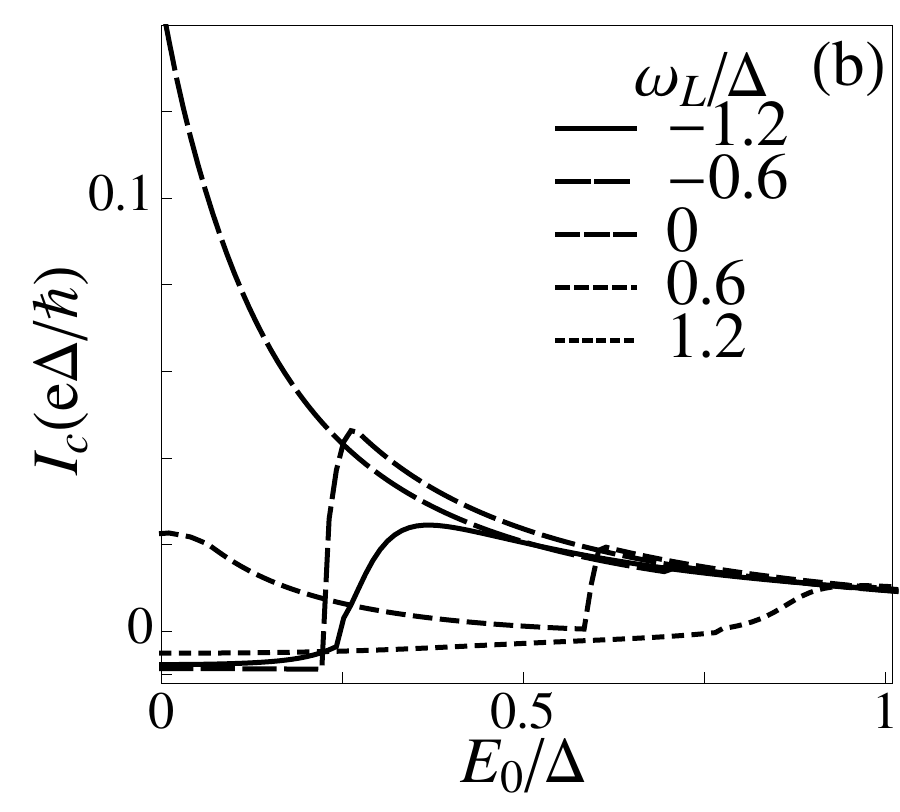}}\vspace{-0.0cm}
\caption{Critical current as a function of $E_0$. The parameters in (a) corresponds to Fig. \ref{figABS} (c) for different exchange couplings at $\vartheta=\pi/4$ and $\omega_L=\Delta/2$.  In (b), $v_s=0.3\Delta$ and $\vartheta=\pi/4$. In both panels $\Gamma=\Delta/10$ and $k_BT=10^{-4}\Delta$.}
\label{figCCa}
\end{centering}
\end{figure}

Figure \ref{figCCa} shows the critical current as a function of the energy level at $ \omega_L=\Delta/2 $ and different couplings $ v_s $ in panel (a). For small $ E_0 $, the state of the junction undergoes a $ \pi $ to $ 0 $ transition by increasing $ v_s $ and finally goes back to the $ \pi $ state. If the junction is in a $ \pi $ state at $ E_0 $, a transition is driven to the $ 0 $ state by increasing $ E_0 $ for all values of the exchange coupling shown in panel (a).   
The critical current at $ v_s=0.1\Delta $ in the range of $ E_0=0 $ to $ E_0=\Delta/2 $ corresponds to the maximal current of each current phase relations shown in Fig. \ref{figCPRvb} (a). The sharp increase of the critical current occurs due to the shift of Andreev states across the Fermi energy within a small parameter range of $ E_0 $. In panel (b), $ v_s=0.3\Delta $ and the critical current is depicted for different values of $ \omega_L $. Again, by increasing $ E_0 $, the junction undergoes a $ \pi $ to $ 0 $ transition at $ \omega_L=0 $ and $ \omega_L = \pm 1.2 \Delta $. In addition, the junction also exhibits multiple transitions for $ E_0 \lesssim \Delta/2 $ as $ \omega_L $ is increased.

\begin{figure}[!t]
\begin{centering}
\subfigure{\hspace{-0.2cm}
\includegraphics[width=0.5\columnwidth]{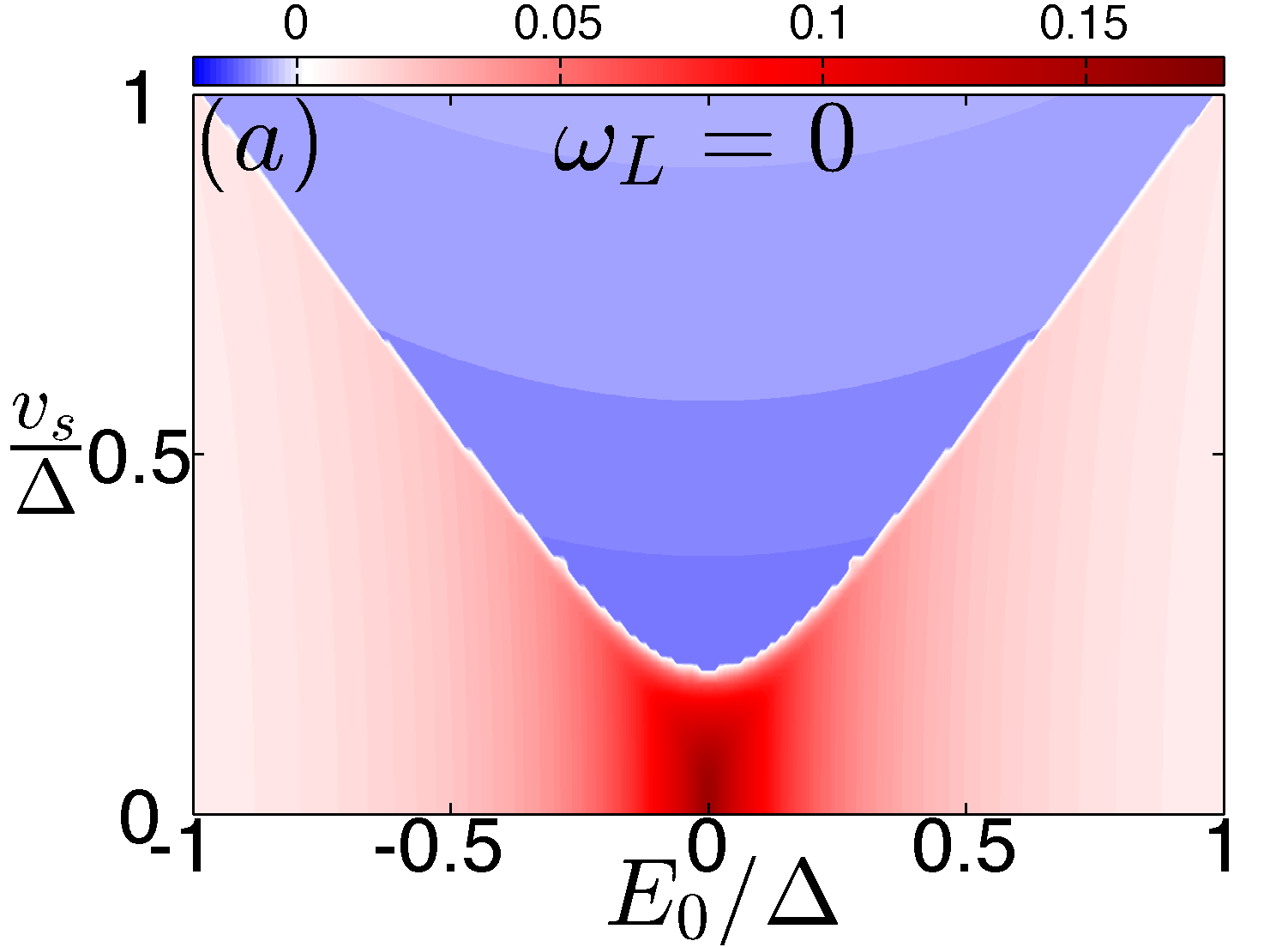}}\vspace{-0.2cm}
\subfigure{\hspace{-0.2cm}
\includegraphics[width=0.5\columnwidth]{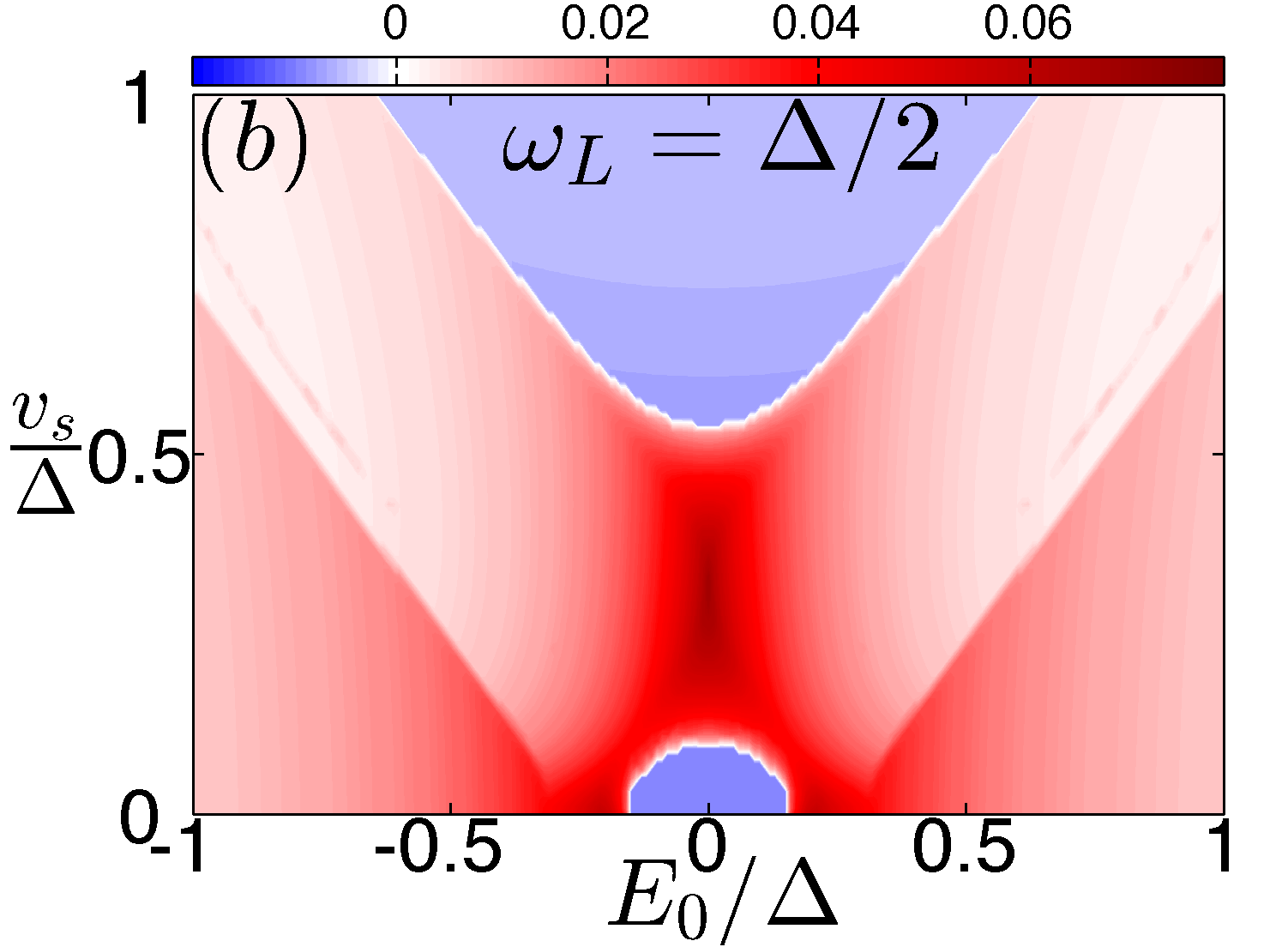}}\vspace{-0.0cm}
\subfigure{\hspace{-0.2cm}
\includegraphics[width=0.5\columnwidth]{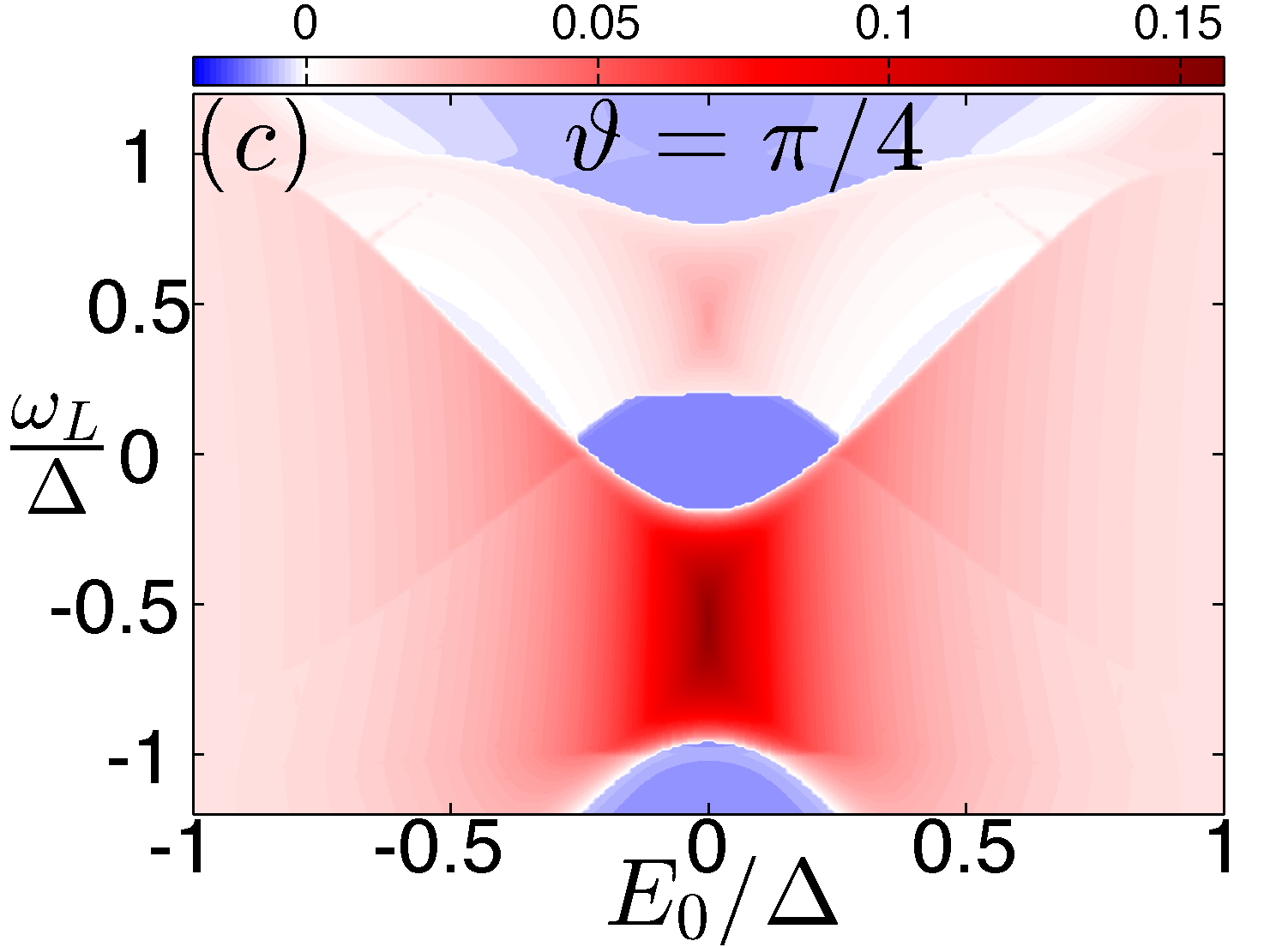}}\vspace{-0.0cm}
\subfigure{\hspace{-0.2cm}
\includegraphics[width=0.5\columnwidth]{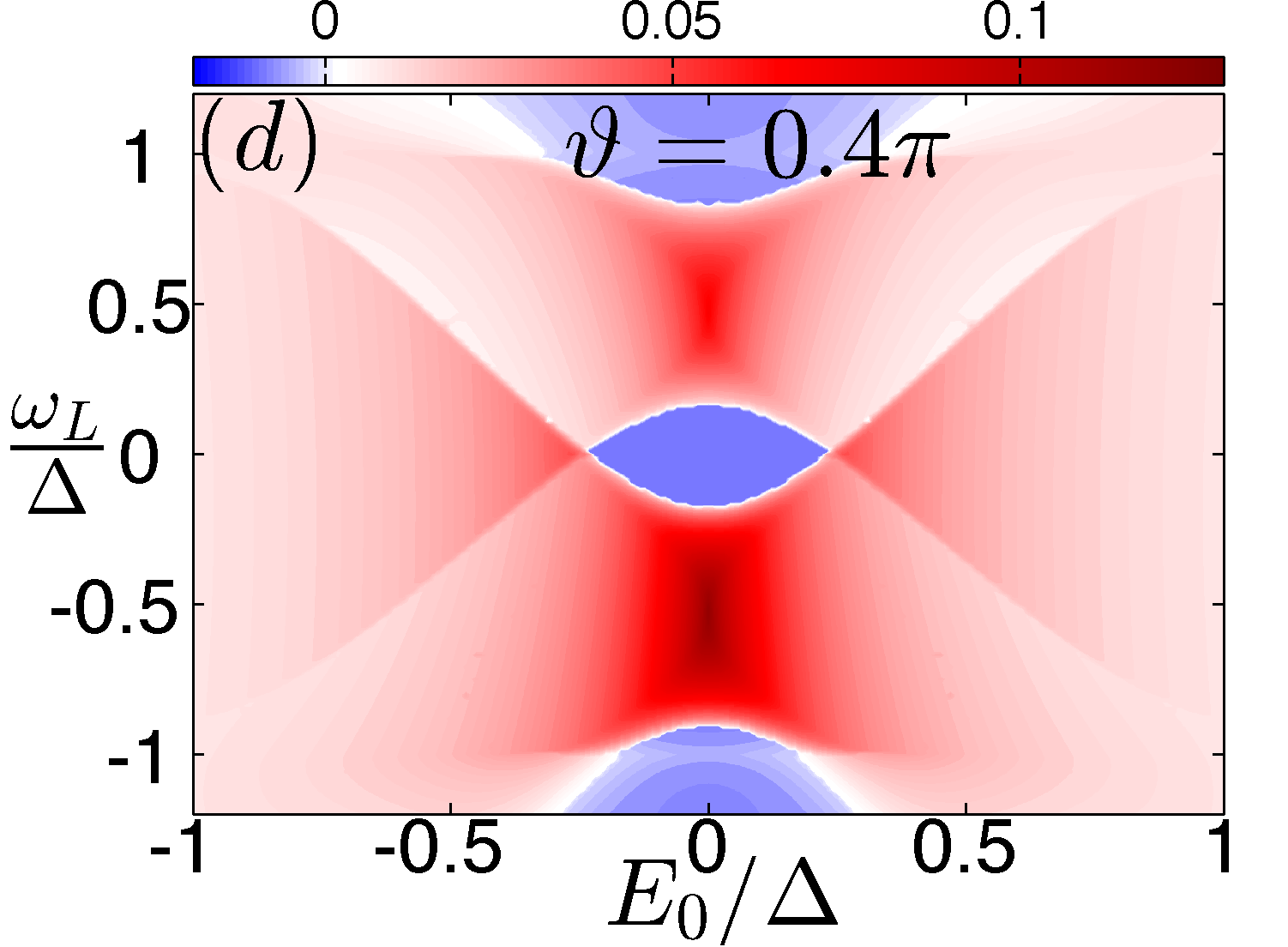}}\vspace{-0.0cm}
\caption{(Color online) Critical current $I_c$ in units of $e\Delta/\hbar$ as a function of $E_0$ and $v_s$ at  $\vartheta=\pi/4$ (panel (a) and (b)). In (a) $\omega_L=0$ and in (b)  $\omega_L=\Delta/2$. In (c) and (d), the critical current is shown as a function of $E_0$ and $\omega_L$ at $v_s=0.3\Delta$. The angle $\vartheta$ varies from  $\vartheta=\pi/4$ in (c) to $\vartheta=0.4\pi$ in (d). The other parameters are $\Gamma=\Gamma_L=\Gamma_R=0.1\Delta$ and $k_BT = 10^{-4}\Delta$. } 
\label{figCCda}
\end{centering}
\end{figure}

Contour plots of the critical current are depicted in Fig. \ref{figCCda}. In all panels, we assume a symmetric coupling between the quantum dot and the leads and calculate the critical current in the low temperature limit. In (a) and (b), the critical current is plotted as a function of $ E_0 $ and $ v_s $ at $ \vartheta=\pi/4 $ and  the magnetic field is set to $ \omega_L=0 $ in (a) and $ \Delta/2 $ in (b). In panel (a), the junction exhibits a $ 0 $ to $ \pi $ transition due to an increase of the exchange coupling, $ v_s $. In this case, the contribution to the current of the Andreev states does not depend on $ \vartheta $ since the magnetic field is zero. The Andreev states intersect the Fermi function at $ v_s = \pm \sqrt{4\Gamma^2\mathrm{cos}(\varphi_{\mathrm{max}}/2)^2+E_0^2} $, with the phase $ \varphi_{\mathrm{max}} $ corresponding to the phase at the critical current. At these values of the exchange coupling, the junction exhibits the transition. If $ v_s \ge  2 \Gamma \vert \mathrm{cos}(\varphi_{max}/2) \vert $, the junctions also exhibits $ \pi $ to $ 0 $ transitions as a result of increasing $ E_0 $.  In panel (b), the magnetic field is set to $ \omega_L=\Delta/2 $ and the Andreev states are degenerate due to the spin-flip scattering. The parameters in this panel correspond to the critical current shown in Fig. \ref{figCCa} (a). Transitions are driven by either increasing $ E_0 $ or $ v_s $. Due to the redistribution of the occupied states at finite $ \omega_L $, the critical current exhibits an additional step which is due to a shift of the Andreev states across the Fermi energy.  In panel (c) and (d), the critical current is shown as a function of $ E_0 $ and $ \omega_L $ at $ v_s=0.3\Delta $ and $ \vartheta $ changes from $ \vartheta=\pi/4 $ to $ \vartheta=0.4\pi $. Due to the spin-flip term, the critical current shows a mirror structure around $ \omega_L=0 $ with different intensities for $ \omega_L \rightarrow -\omega_L$. As discussed in Sec. \ref{subsec:DOS}, the Andreev states are symmetric with respect to $ \omega_L \rightarrow -\omega_L$ at $ \vartheta=0.5\pi $. Therefore, the critical current shows a symmetric behaviour as a function of $ \omega_L $ since $ \vartheta $ is close to $ 0.5\pi $. For small $ E_0 $, the junctions exhibits multiple transitions as a function of the magnetic field.

\section{Conclusions}\label{sec: conclusions}
We have studied the electronic transport in the low temperature limit of a Josephson junction consisting of two superconducting leads and a quantum dot connected to a molecular magnet. The exchange interaction lifts the spin degeneracy of the energy level of the dot as well as of the Andreev states. The current-phase relation shows a strong modification when an Andreev state intersects the Fermi energy and the Josephson junction can be driven into the $ \pi $ state by the exchange interaction (Fig. \ref{figCPRABS}(a)). A magnetic field applied to the central region induces a Zeeman energy of the electrons on the quantum dot and a precession of the molecular magnet's magnetisation. As a result, an electron on the dot can undergo spin-flip scattering and the occupation of the Andreev states is redistributed by the magnetic field. The redistribution depends on the orientation between the magnetic field and magnetisation of the molecular magnet and can lead to an increase in the current-phase relation (Fig. \ref{figCPRABS}(c)) when the electron-like quasiparticles are scattered into states below the Fermi energy carrying the current in positive direction. The critical current in Fig. \ref{figCCda} shows the possibility to observe multiple transitions as a function of a gate voltage which changes the energy level of the dot. Multiple transitions are also obtained if the magnetic field is swept around zero. However, in a more realistic system the dependence of the superconducting gap on the magnetic field must be taken into account in a self-consistent way.  

In principle, the results of the paper can be used in two ways. First, the measurement of the physical magnitudes of either the energy of Andreev levels, the current-phase relation or the critical current, allows to reveal the internal structure of the Josephson junction. Different parameters are possible to extract from measurements such as the strength of the exchange interaction. Second, the presence of the molecular magnet offers the possibility to manipulate the current through the junction by changing the parameters such as the energy level of the dot or the magnetic field. This manipulation strongly affects the Andreev states which allows possible application in quantum computation. \cite{Chtchelkatchev:2003ji,Michelsen:2008wv}

\section*{Acknowledgments}
This work was supported by Deutsche Forschungsgemeinschaft and SFB 767. 


\begin{thebibliography}{99}
\bibitem{Rocha:2005ep} A. R. Rocha, V. M. Garc{\'i}a-su{\'a}rez, S. W. Bailey, C. J. Lambert, J. Ferrer, and S. Sanvito,
Nature Materials \textbf{4}, 335 (2005)
\bibitem{Wolf:2001fu}  S. A. Wolf, Science \textbf{294}, 1488 (2001).
\bibitem{Bogani:2008tc}  L. Bogani and W. L. Wernsdorfer, Nature Materials \textbf{7}, 179 (2008).
\bibitem{MolecularSpintronic:2006de} S. Sanvito and A. R. Rocha, J. Comput. Theor. Nanosci. \textbf{3}, 624 (2006).
\bibitem{Leuenberger:478346} M. N. Leuenberger and D. Loss, Nature \textbf{410}, 789 (2001).
\bibitem{Timm:2012cf} C. Timm and M. Di Ventra, Phys. Rev. B \textbf{86}, 104427 (2012).
\bibitem{Ardavan:2007ci} A. Ardavan, O. Rival, J. J. L. Morton, S. J. Blundell, A. M. Tyryshkin, G. A. Timco, and R. E. P. Winpenny, Phys. Rev. Lett. \textbf{98}, 057201 (2007).
\bibitem{Wernsdorfer:1999tx} W. Wernsdorfer and R. Sessoli, Science \textbf{284}, 133 (1999).
\bibitem{Wernsdorfer:2002jx} W. Wernsdorfer, M. Soler, G. Christou, and D. N. Hendrickson, J. Appl. Phys. \textbf{91}, 7164 (2002).
\bibitem{Christou:2000vo} G. Christou, D. Gatteschi, D. N. Hendrickson, and R. Sessoli, MRS Bulletin \textbf{25}, 66 (2000).
\bibitem{Gatteschi:2003uu}  D. Gatteschi and R. Sessoli, Angewandte Chemie International Edition \textbf{42}, 268 (2003).
\bibitem{Grose:2008ev}  J. E. Grose, E. S. Tam, C. Timm, M. Scheloske, B. Ulgut, J. J. Parks, H. D. Abru{\~n}a, W. Harneit, and D. C. Ralph, Nature Materials \textbf{7}, 884 (2008).
\bibitem{Jo:2006ck} M.-H. Jo, J. E. Grose, K. Baheti, M. M. Deshmukh, J. J. Sokol, E. M. Rumberger, D. N. Hendrickson, J. R. Long, H. Park, and D. C. Ralph, Nano Lett. \textbf{6}, 2014 (2006).
\bibitem{Heersche:2006cs} H. B. Heersche, Z. de Groot, J. A. Folk, H. S. J. van der Zant, C. Romeike, M. R. Wegewijs, L. Zobbi, D. Barreca, E. Tondello, and A. Cornia, Phys. Rev. Lett. \textbf{96}, 206801 (2006).
\bibitem{Henderson:2007vi}  J. J. Henderson, C. M. Ramsey, E. Del Barco, A. Mishra, and G. Christou, J. Appl. Phys. \textbf{101}, 09E102 (2007).
\bibitem{Haque:2011we} F. Haque, M. Langhirt, E. Del Barco, T. Taguchi, and G. Christou, J. Appl. Phys. \textbf{109}, 07B112 (2011).
\bibitem{Zyazin:2012gc} A. S. Zyazin, J. W. G. van den Berg, E. A. Osorio, H. S. J. van der Zant, N. P. Konstantinidis, M. Leijnse, M. R. Wegewijs, F. May, W. Hofstetter, C. Danieli, and A. Cornia, Nano Lett. \textbf{10}, 3307 (2012).
\bibitem{Burzuri:2012gb} E. Burzur{\'i}, A. S. Zyazin, A. Cornia, and H. S. J. van der Zant, Phys. Rev. Lett. \textbf{109}, 147203 (2012).
\bibitem{Urdampilleta:2011ii}  M. Urdampilleta, Nature Materials \textbf{10}, 502 (2011).
\bibitem{Candini:2011ul} A. Candini, S. Klyatskaya, M. Ruben, W. Wernsdorfer, and M. Affronte, Nano Lett. \textbf{11}, 2634 (2011).
\bibitem{Kasumov:2005dn} A. Y. Kasumov, K. Tsukagoshi, M. Kawamura, T. Kobayashi, Y. Aoyagi, K. Senba, T. Kodama, H. Nishikawa, I. Ikemoto, K. Kikuchi, V. T. Volkov, Y. A. Kasumov, R. Deblock, S. Gu{\'e}ron, and H. Bouchiat, Phys. Rev. B \textbf{72}, 033414 (2005).
\bibitem{Winkelmann:2009vc} C. B. Winkelmann, N. Roch, W. Wernsdorfer, V. Bouchiat, and F. Balestro, Nat. Phys. \textbf{5}, 876 (2009).
\bibitem{Clarke:2008gi} J. Clarke and F. K. Wilhelm, Nature \textbf{453}, 1031 (2008).
\bibitem{Wendin:2007da} G. Wendin and V. S. Shumeiko, Low Temp. Phys. \textbf{33}, 724 (2007).
\bibitem{Zazunov:2003jm} A. Zazunov, V. S. Shumeiko, E. N. Bratus, J. Lantz, and G. Wendin, Phys. Rev. Lett. \textbf{90}, 087003 (2003).
\bibitem{Zazunov:2005ec}  A. Zazunov, V. S. Shumeiko, G. Wendin, and E. N. Bratus, Phys. Rev. B \textbf{71}, 214505 (2005).
\bibitem{Chtchelkatchev:2003ji}  N. M. Chtchelkatchev and Y. V. Nazarov, Phys. Rev. Lett. \textbf{90}, 226806 (2003).
\bibitem{Zgirski:2011dx} M. Zgirski, L. Bretheau, Q. Le Masne, H. Pothier, D. Esteve, and C. Urbina, Phys. Rev. Lett. \textbf{106}, 257003 (2011).
\bibitem{Michelsen:2008wv}  J. Michelsen, V. S. Shumeiko, and G. Wendin, Phys. Rev. B \textbf{77}, 184506 (2008).
\bibitem{Pillet:2010ds} J.-D. Pillet, C. H. L. Quay, P. Morfin, C. Bena, A. L. Yeyati, and P. Joyez, Nat. Phys. \textbf{6}, 965 (2010).
\bibitem{DellaRocca:2007ua} M. L. Della Rocca, M. Chauvin, B. Huard, H. Pothier, D. Esteve, and C. Urbina, Phys. Rev. Lett. \textbf{99}, 127005 (2007).
\bibitem{Bulaevskii:1977uj} L. N. Bulaevskii, V. V. Kuzii, and A. A. Sobyanin, JETP Lett \textbf{25}, 290 (1977).
\bibitem{vanDam:2006fj} J. A. van Dam, Y. V. Nazarov, E. P. A. M. Bakkers, S. De Franceschi, and L. P. Kouwenhoven, Nature \textbf{442}, 667 (2006).
\bibitem{Ryazanov:2001jp} V. V. Ryazanov, V. A. Oboznov, A. Y. Rusanov, A. V. Veretennikov, A. A. Golubov, and J. Aarts, Phys. Rev. Lett. \textbf{86}, 2427 (2001).
\bibitem{Kontos:2002hj} T. Kontos, M. Aprili, J. Lesueur, F. Gen{\^e}t, B. Stephanidis, and R. Boursier, Phys. Rev. Lett. \textbf{89}, 137007 (2002).
\bibitem{Benjamin:2007fz} C. Benjamin, T. Jonckheere, A. Zazunov, and T. Martin, Eur. Phys. J. B \textbf{57}, 279 (2007).
\bibitem{Lee:2008ka} M. Lee, T. Jonckheere, and T. Martin, Phys. Rev. Lett. \textbf{101}, 146804 (2008).
\bibitem{Sadovskyy:2011ge} I. A. Sadovskyy, D. Chevallier, T. Jonckheere, M. Lee, S. Kawabata, and T. Martin, Phys. Rev. B \textbf{84}, 184513 (2011).
\bibitem{Teber:2010bo} S. Teber, C. Holmqvist, and M. Fogelstr\"om, Phys. Rev. B \textbf{81}, 174503 (2010).
\bibitem{Holmqvist:2011bv} C. Holmqvist, S. Teber, and M. Fogelstr\"om, Phys. Rev. B \textbf{83}, 104521 (2011).
\bibitem{Holmqvist:2012jz} C. Holmqvist, W. Belzig, and M. Fogelstr\"om, Phys. Rev. B \textbf{86}, 054519 (2012).
\bibitem{Holmqvist:2013uj} C. Holmqvist, M. Fogelstr\"om, and W. Belzig, arXiv:1302.7107 (2013).
\bibitem{Nussinov:2005id} Z. Nussinov, A. Shnirman, D. P. Arovas, A. V. Balatsky, and J. X. Zhu, Phys. Rev. B \textbf{71}, 214520 (2005).
\bibitem{Zhu:2004dn} J. X. Zhu, Z. Nussinov, A. Shnirman, and A. V. Balatsky, Phys. Rev. Lett. \textbf{92}, 107001 (2004).
\bibitem{Roch:2011ht} N. Roch, R. Vincent, F. Elste, W. Harneit, W. Wernsdorfer, C. Timm, and F. Balestro, Phys. Rev. B \textbf{83}, 081407 (2011).
\bibitem{Tserkovnyak:2005zz} Y. Tserkovnyak, A. Brataas, G. E. Bauer, and B.I. Halperin, Rev. Mod. Phys. \textbf{77}, 1375 (2005). 
\bibitem{Kittel:1948ur} C. Kittel, Physical Review \textbf{73}, 155 (1948).
\bibitem{Bell:2008gh}  C. Bell, S. Milikisyants, M. Huber, and J. Aarts, Phys. Rev. Lett. \textbf{100}, 047002 (2008).
\bibitem{Eilenberger:1968bb}  G. Eilenberger, Z. Physik \textbf{214}, 195 (1968).
\bibitem{Larkin:1969wa} A. I. Larkin and Y. N. Ovchinnikov, Sov. Phys. JETP \textbf{28}, 1200 (1969); Zh. Eksp. Teor. Fiz. \textbf{55}, (1968).
\bibitem{Serene:1983vc} J. W. Serene and D. Rainer, Physics Reports \textbf{101}, 221 (1983).
\bibitem{Cuevas:2001km} J. C. Cuevas and M. Fogelstr\"om, Phys. Rev. B \textbf{64}, 104502 (2001).
\bibitem{Kopnin:2009wt} N. B. Kopnin, Theory of non equilibrium superconductivity, Oxford University Press (2009)
\bibitem{Jauho:1994te} A. P. Jauho, N. S. Wingreen, and Y. Meir, Phys. Rev. B \textbf{50}, 5528 (1994).
\bibitem{Cuevas:TshMQ34R} J. C. Cuevas and E. Scheer, Molecular Electronics: an introduction to theory and experiment, World Scientific (2010).
\bibitem{Beenakker:2001wo}  C. Beenakker and H. Van Houten, In: Single-Electron Tunneling and Mesoscopic Devices, edited by H. Koch and H. L\"ubbig (Springer, Berlin, 1992): pp. 175-179 (2001).
\bibitem{Golubov:2004zz}  A. A. Golubov, M. Y. Kupriyanov, and E. Il'ichev, Rev. Mod. Phys. \textbf{76}, 411 (2004).
\end{thebibliography}
\end{document}